\documentclass[twocolumn]{aastex62}

\usepackage{amsmath}
\usepackage{nicefrac}
\usepackage{tensor}

\newcommand{\harm}{{\tt harm}}
\newcommand{\bhlight}{{\tt bhlight}}

\newcommand{\ug}{u_{\rm g}}

\newcommand{\jnua}{j_{\nu}^{\rm a}}

\graphicspath{{./}{figures/}}

\submitjournal{ApJS}

\shorttitle{MOCMC: Method of Characteristics Moment Closure}
\shortauthors{Ryan and Dolence}

\begin{document}

\title{MOCMC: Method of Characteristics Moment Closure, a Numerical Method for Covariant Radiation Magnetohydrodynamics}

\author{Benjamin R. Ryan}
\affiliation{CCS-2, Los Alamos National Laboratory, P.O. Box 1663, Los Alamos, NM 87545, USA}
\affiliation{Center for Theoretical Astrophysics, Los Alamos National Laboratory, Los Alamos, NM, USA}

\author{Joshua C. Dolence}
\affiliation{CCS-2, Los Alamos National Laboratory, P.O. Box 1663, Los Alamos, NM 87545, USA}
\affiliation{Center for Theoretical Astrophysics, Los Alamos National Laboratory, Los Alamos, NM, USA}

\begin{abstract}

We present a conservative numerical method for radiation magnetohydrodynamics with frequency-dependent full transport in stationary spacetimes. This method is stable and accurate for both large and small optical depths and radiation pressures. The radiation stress-energy tensor is evolved in flux-conservative form, and closed with a swarm of samples that each transport a multigroup representation of the invariant specific intensity along a null geodesic. In each zone, the enclosed samples are used to efficiently construct a Delaunay triangulation of the unit sphere in the comoving frame, which in turn is used to calculate the Eddington tensor, average source terms, and adaptively refine the sample swarm. Radiation four-fources are evaluated in the moment sector in a semi-implicit fashion. The radiative transfer equation is solved in invariant form deterministically for each sample.  Since each sample carries a discrete representation of the full spectrum, the cost of evaluating the transport operator is independent of the number of frequency groups, representing a significant reduction of algorithmic complexity for transport in frequency dependent problems. The major approximation we make in this work is performing scattering in an angle-averaged way, with Compton scattering further approximated by the Kompaneets equation. Despite relying on particles to solve the radiative transfer equation, the scheme is efficient and stable for both large optical depths and small ratios of gas to radiation pressure. Local adaptivity in samples also makes this scheme more amenable to nonuniform meshes than a traditional Monte Carlo method. We describe the method and present results on a suite of test problems. We find that MOCMC converges at least as $\sim N^{-1}$, rather than the canonical Monte Carlo $N^{-1/2}$, where $N$ is the number of samples per zone. Isotropic one-zone problems have no shot noise at all. On several problems we demonstrate substantial improvement over Eddington and M1 closures and gray opacities.

\end{abstract}

\section{Introduction} \label{sec:intro}

In astrophysics, radiation often plays an important role in transporting energy and momentum. Accretion disks around neutron stars and black holes are subject to perturbative radiative cooling (\citealt{Esin+1997}, \citealt{Ryan+2017}, \citealt{Sadowski+2017}) at the lowest accretion rates, efficient, local radiative losses (\citealt{ShakuraSunyaev1973}, \citealt{Jiang+2019}) near the Eddington limit, and significant photon trapping and dominant radiation pressures at super-Eddington rates (\citealt{Abramowicz+1988}, \citealt{Sadowski+2014}, \citealt{Jiang+2014b}, \citealt{McKinney+2014}). At least some core-collapse supernova explosions are probably driven by energy transfer via neutrinos (\citealt{Burrows+1995}, \citealt{Janka+2007}, \citealt{Vartanyan+2019}). The composition of ejecta from merging binary neutron stars, probably crucial for setting the color of kilonovae (\citealt{Metzger+2010}), is affected by neutrino fluxes, as the neutrinos also transport lepton number (\citealt{Surman+2008}, \citealt{Wanajo+2014}, \citealt{Foucart+2015}, \citealt{Richers+2015}, \citealt{Foucart+2016}, \citealt{Miller+2019a}, \citealt{Miller+2019b}). The envelopes of high mass stars can be radiation pressure-supported (\citealt{Paxton+2013}, \citealt{Jiang+2018}).

In such systems magnetohydrodynamic (MHD) turbulence also often plays an important role.
In particular, MHD turbulence may dominate angular momentum transport in black hole accretion disks (\citealt{BalbusHawley1991}, \citealt{Hawley+1995}).
Here, and in other relativistic flows, the fluid sound speed is approximately the speed of light. Therefore, coupling time-dependent radiative transfer to time-dependent MHD turbulence is required for accurately modeling these systems from first principles.

Solving the equations of radiation hydrodynamics presents significant difficulties. In particular, the specific intensity is in general a function of spatial location, frequency, and direction. This leads to a high dimensional integration (3 space + 3 momentum + 1 time). For global turbulent flows, where the the requirement for resolving the flow locally can impose a large minimum number of grid zones and irregular spatial grids, a lack of computational efficiency can be prohibitive. Additional conceptual difficulties can also arise when considering radiation transport in curvilinear coordinates and/or with large Lorentz factors. Timescales for energy and momentum exchange between fluid and radiation can also be short compared to global dynamical timescales. 

For large optical depths $\tau$, the Eddington approximation along with averaging opacities over the Planck function is a straightforward, effective approach. However, there is not a clear hierchical process by which this approach can be extended out of the optically thick regime. The M1 family of closures, in which the entire four-momentum rather than just the comoving radiation energy density is used to close the second moment of the radiation, is frequently adopted (\citealt{Minerbo1978}, \citealt{Levermore1984}, \citealt{Scheck+2006}, \citealt{Sadowski+2013}, \citealt{McKinney+2014}, \citealt{Foucart+2015}, \citealt{Roberts+2016}, \citealt{Skinner+2019}). However, while it recovers the optically thick isotropic limit, M1 can represent only a specific case of optically thin transport in which the radiation is isotropic in the rest frame of some timelike observer. Truncated moment methods without a separate solution of the radiative transfer equation will in general be forced to make assumptions about the structure of the radiation distribution function, and such a closure that is accurate across problems of interest in astrophysical radiation transport is unknown.

A classic approach to radiation transport is the Monte Carlo method (e.g.\ \citealt{FleckCummings1971}, \citealt{Pozdnyakov+1983}, \citealt{Dolence+2009}, \citealt{Abdikamalov+2012}, \citealt{SchnittmanKrolik2013}, \citealt{Wollaeger+2013}, \citealt{RothKasen2015}, \citealt{Ryan+2015}, \citealt{Wollaber2016}). In Monte Carlo methods, the radiation is randomly sampled and these samples undergo transport and interactions as if they were individual particles. Advantages of this method include the unbiased nature of Monte Carlo sampling, simple extension to frequency dependence (even to a continuous energy approach), and the simplicity of interpreting the method as individual physical interactions. However, the method converges with the number of samples $N$ only as $N^{-1/2}$, although there is no scaling with number of dimensions. This error will be roughly divided by $\beta_{\rm r}$, the ratio of gas to radiation pressure, before being felt by the fluid; reducing sampling error sufficiently when radiation pressures are large is generally not practical. The requirement of resolving interactions in an unbiased manner is also onerous when optical depths are large.

Monte Carlo methods can be extended in several ways. Implicit Monte Carlo (IMC; \citealt{FleckCummings1971}, \citealt{Wollaber2016} for a review) linearizes the source terms and converts a fraction of emission and absorption events into effective scatterings. To be precise, this method is semi-implicit; the linearization can lead to unphysical behavior like violation of a maximum principle (\citealt{LarsenMercier1987}, although see e.g.\ \citealt{ClevelandWollaber2018}). Inelastic scattering (e.g.\ Compton scattering) is also not amenable to the effective scattering approach, and poses a particular challenge (\citealt{Densmore+2010}). Additionally, the method can still experience significant slowdowns due to large numbers of effective scatterings in optically thick regions. Random walk methods (\citealt{FleckCanfield1984}) alleviate this by updating particle positions according to a probabilistic solution of a diffusion equation. In a similar vein, Discrete Diffusion Monte Carlo (\citealt{Densmore+2007}) changes the character of optically thick zones and can be combined with IMC (\citealt{Abdikamalov+2012}, \citealt{Densmore+2012}, \citealt{Wollaeger+2013}), but this requires a heuristic choice of interface between optically thin and thick regions. These modifications do little to enhance stability when the radiation pressure is large.

Another standard approach, in which the intensity is discretized into rays on a per-zone basis, is the method of discrete ordinates (e.g.\ \citealt{Liebendorfer+2004} \citealt{HubenyBurrows2007}, \citealt{Ott+2008}, \citealt{Davis+2012}, \citealt{OhsugaTakahashi2016}, \citealt{Nagakura+2017}). It is straightforward to make this method stable and efficient for both optically thick and radiation pressure-dominated flows, as the transport equation is solved in a deterministic fashion. However, these methods suffer from ray effects in optically thin regions; the truncation error is highly anisotropic (\citealt{Castor2004}, \citealt{Zhu+2015}). Additionally, when these methods are made frequency-dependent, a separate transport update is required for every frequency element of every angle, producing a scheme that is often prohibitively expensive.

Another approach to transport intended to yield good stability properties in radiation hydrodynamics problems is the quasidiffusion or variable Eddington tensor (VET) scheme (\citealt{Goldin1964}, \citealt{Takeuchi1971}, \citealt{Stone+1992}, \citealt{Davis+2012}, \citealt{Jiang+2012}). Here, the zeroth and first moments of the radiation are evolved, but the Eddington tensor is evaluated with a separate solution of the full transport equation. \cite{Foucart2018} adopted a similar approach for relativistic systems, using Monte Carlo rather than discrete ordinates to evaluate the Eddington tensor. The VET allows for full transport while evaluating radiation-fluid interactions for only a few degrees of freedom (the four-momentum), which can greatly simplify semi-implicit methods often required for stability.

Here we introduce a method that we term Method of Characteristics Moment Closure (MOCMC). As in VET approaches, we evolve the frequency-integrated (gray) moments of the specific intensity, closed by a solution of the full transport equation. However, instead of using a Monte Carlo or discrete ordinates method to obtain the transport solution, we adopt a Method of Characteristics approach (\citealt{Askew1972}, \citealt{PandyaAdams2009}, \citealt{Park+2019}, \citealt{Hammer+2019}). In our method, the radiation is discretized into samples with different positions and directions, with each sample carrying an array of specific intensities discretized in frequency. These samples move along characteristics, and the radiative transfer equation is solved in a deterministic way. The samples in each zone are used to reconstruct momentum space in that zone (we use the Delaunay triangulation of the unit sphere in the comoving frame). The reconstructed intensity is used to evaluate the pressure tensor and frequency- and angle-average source terms in order to close the moment equations in a VET fashion.

Our approach has several advantages. Specific intensities are integrated directly along geodesics; the transport process itself is essentially free of spatial discretization errors. The deterministic approach to source terms puts a limit on the computational cost of the scheme (unlike e.g. probabilistic interaction-by-interaction scattering of Monte Carlo particles in optically thick media). Our finite volume interpretation frees us from issues with positivity and oscillations encountered in spectral methods (although see \citealt{McClarrenHauck2010}, \citealt{Radice+2013}). No representation of a conserved four-momentum is evaluated from the samples; characteristics crossing zones does not lead to intrinsic noise, and allows for significant freedom in dynamically refining or derefining characteristic resolution. This also frees us from having to reconcile different four-momentum representations in the radiation moments and the samples. Absent plasma dispersion effects, photons of all frequencies share geodesics, so we can efficiently advect multiple frequency bins with a single push along a characteristic i.e.\ each resolution element carries an array of specific intensities at different frequencies. Relaxing to the asymptotic diffusion limit is straightforward in the moment sector. Finally, our scheme employs relativistic invariants in the samples, leading to conceptual simplicity.

We begin with a description of the governing equations of radiation magnetohydrodynamics in Section \ref{sec:eqns}. We then describe our numerical implementation in Section \ref{sec:method}. We present a suite of tests in Section \ref{sec:tests}, in which we also compare our scheme directly to gray moment methods in several cases. We conclude in Section \ref{sec:conclusion}. 

\section{Equations of Radiation Hydrodynamics}
\label{sec:eqns}

Throughout this work, we use parentheses to denote indices in an orthonormal tetrad frame. Indices without parentheses indicate the coordinate frame. Greek letters index spacetime $(0,1,2,3)$, while Latin letters index space $(1,2,3)$. We adopt units such that $c=1$.

We consider the equations of radiation magneto\-hydrodynamics (RMHD), including the full transport equation, written in covariant form. Unlike in mixed-frame approaches (\citealt{MihalasKlein1982}, \citealt{MihalasMihalas1984}, \citealt{Krumholz+2007}, \citealt{Jiang+2014a}, \citealt{Skinner+2019}), here it is not necessary to expand the equations in powers of $v/c$ in order to relate the radiation and fluid. The equations are valid in any frame.

The divergence of the total stress-energy tensor is zero:
\begin{align}
\tensor{T}{^{\mu}_{\nu;\mu}} + \tensor{R}{^{\mu}_{\nu;\mu}} = 0
\end{align}
where $\tensor{T}{^{\mu}_{\nu}}$ is the stress-energy of the MHD fluid and $\tensor{R}{^{\mu}_{\nu}}$ is the stress-energy of the radiation. Rewriting this expression, evidently the fluid
 and radiation interact via exchange of four-momentum:
\begin{align}
\tensor{T}{^{\mu}_{\nu;\mu}} = -\tensor{R}{^{\mu}_{\nu;\mu}}.
\label{eqn:gasrad}
\end{align}

We first separately consider the evolution of fluid and radiation, and then describe the radiation source terms (emission, absorption, and scattering) that lead to four-momentum exchange between these two components.

\subsection{Magnetohydrodynamics}

The governing equations of covariant magnetohydrodynamics, written in conservative form in a coordinate basis, are (\citealt{Anile1989}, \citealt{Komissarov1999}, \citealt{Gammie+2003}) conservation of particle number
\begin{align}
\left( \sqrt{-g} \rho u^{t}\right)_{,t} = -\left( \sqrt{-g} \rho u^{i}\right)_{,i}
\end{align}
conservation of four-momentum
\begin{align}
\left( \sqrt{-g} \tensor{T}{^{t}_{\nu}}\right)_{,t} = -\left( \sqrt{-g} \tensor{T}{^{i}_{\nu}}\right)_{,i} + \sqrt{-g} \tensor{T}{^{\kappa}_{\lambda}} \tensor{\Gamma}{^{\lambda}_{\nu \kappa}}
\end{align}
conservation of magnetic flux
\begin{align}
\left( \sqrt{-g} B^i\right)_{,t}= -\left( \sqrt{-g} \left[b^ju^i - b^iu^j\right]\right)_{,j}
\end{align}
and the no-monopoles constraint
\begin{align}
\left(\sqrt{-g} B^i \right)_{,i} = 0.
\label{eqn:divb}
\end{align}
The MHD stress-energy tensor is
\begin{align}
\tensor{T}{^{\mu}_{\nu}} &= \left( \rho + \ug + P_{\rm g} + b^2 \right)u^{\mu} u_{\nu}  \\
&+\left(P_{\rm g} + \frac{b^2}{2} \right)\tensor{\delta}{^{\mu}_{\nu}} - b^{\mu}b_{\nu}.
\end{align}
In the above, $\rho$ is the fluid rest-mass density, $u^{\mu}$ is the fluid four-velocity, $B^i$ is the magnetic field three-vector, $b^{\mu}$ is the magnetic field four-vector, and $\sqrt{-g}$ is the determinant of the metric. These equations require an equation of state; throughout this work we adopt $P_{\rm g} = (\gamma - 1) \ug$, with $\ug$ the gas internal energy density, although introducing more sophisticated equations of state in this framework is conceptually straightforward (e.g.\ \citealt{Miller+2019a}).

\subsection{Radiation}
The equation of radiation transport, written in invariant form (each quantity in parentheses is invariant), is (\citealt{MihalasMihalas1984})
\begin{widetext}
\begin{align}
\frac{D}{d s}\left( \frac{I_{\nu}}{\nu^3}\right) = \left( \frac{\jnua}{\nu^2} \right) + \left( \frac{j_{\nu}^{\rm s}}{\nu^2} \right) - \left( \nu \alpha_{\nu}^{\rm a} \right)\left( \frac{I_{\nu}}{\nu^3}\right) - \left( \nu \alpha_{\nu}^{\rm s} \right)\left( \frac{I_{\nu}}{\nu^3}\right)
\label{eqn:transfer}
\end{align}
\end{widetext}
where $D/ds$ is the convective derivative in phase space, $s$ is the path length, $\jnua$ is the specific thermal emissivity, $\alpha_{\nu}^{\rm a}$ is the specific absorption coefficient, $j_{\nu}^{\rm s}$ is the effective specific emission coefficient due to scattering (\citealt{MihalasMihalas1984}), and $\alpha_{\nu}^{\rm s}$ is the specific absorption coefficient due to scattering. The specific intensity $I_{\nu}$ is related to the distribution function $f$ as $I_{\nu} = h^4 \nu^3 f / c^2$.
We will return to source terms in Section \ref{sec:fluidradinteractions}; for now it is sufficient to note that, absent interaction terms, Equation \ref{eqn:transfer} preserves the invariant intensity $I_{\nu}/\nu^3$ along characteristics.

Each characteristic is described by a position $x^{\mu}$ and a direction vector $n_{\mu}$. These quantities evolve according to the geodesic equation (note that the coordinate time $t = x^0$),
\begin{align}
\frac{d x^i}{dx^0} &= \frac{n^i}{n^0} \\
\frac{d n_{\mu}}{d x^0} &= -\frac{1}{2n^0}\frac{\partial g^{\nu\lambda}}{\partial x^{\mu}}n_{\nu}n_{\lambda}.
\label{eqn:geodesic}
\end{align}

For Monte Carlo particles, like superphotons in \bhlight{} (\citealt{Ryan+2015}), one would typically use the wavevector $k^{\mu}$ rather than $n^{\mu}$. However, we will discretize $I_{\nu}/\nu^3$ in frequency along each characteristic, so there is no special meaning to the normalization of the direction vector. In fact, $n^{\mu}$ is normalized such that $n_0 = -1$; the frequency $\nu$ of each bin $i$ according to an observer moving with four-velocity $u^{\mu}$ is then
\begin{align}
\nu = -\frac{u^{\mu} n_{\mu} k_0^i }{h}
\end{align}
where $k_0^i$ is an array of timelike components of covariant four-momenta, or for our purposes, frequencies at asymptotic spatial infinity common to all samples\footnote{Such $k_0^i$ would not be available in time-dependent spacetimes, such as near merging compact objects.}. This array defines the range of the frequency discretization. We discretize these frequencies logarithmically; note that $\Delta \log \nu$ is unaffected by frame transformations.

The moments of the radiation field evolve in a similar manner to the MHD stress-energy tensor,
\begin{align}
\left( \sqrt{-g} \tensor{R}{^{t}_{\nu}}\right)_{,t} = -\left( \sqrt{-g} \tensor{R}{^{i}_{\nu}}\right)_{,i} + \sqrt{-g} \tensor{R}{^{\kappa}_{\lambda}} \tensor{\Gamma}{^{\lambda}_{\nu \kappa}},
\end{align}
where $\tensor{R}{^{\mu}_{\nu}}$ is related to the specific intensity $I_{\nu}$ as
\begin{align}
\tensor{R}{^{\mu}_{\nu}} = \frac{1}{h^4}\int \frac{d^3 p}{\sqrt{-g} p^t}p^{\mu}p_{\nu}\left( \frac{I_{\nu}}{\nu^3}\right).
\end{align}
However, unlike in the MHD case where we assumed a thermal, isotropic distribution of particles in the fluid frame, there is no general analytic expression for $R^{i}_{~j}$. In the fluid frame, this uncertainty can be parameterized as (recall that parentheses indicate an orthonormal tetrad) $\tensor{R}{^{(i)}_{(j)}} = -\tensor{R}{^{(0)}_{(0)}}\tensor{\pi}{^{(i)}_{(j)}}$, where $\tensor{\pi}{^{(i)}_{(j)}}$ is the Eddington tensor, which is calculated from the specific intensity:
\begin{align}
\tensor{\pi}{^{(i)}_{(j)}} = \frac{\int d \nu d\Omega I_{\nu} n^{(i)}n_{(j)}}{\int d \nu d\Omega I_{\nu}}.
\label{eqn:eddtensor}
\end{align}
where $d\Omega$ is the differential solid angle in an orthonormal tetrad. Given $\tensor{\pi}{^{(i)}_{(j)}}$ and $\tensor{R}{^0_{\nu}}$, we can construct the entire radiation stress-energy tensor in any frame.

For the special case of very optically thick flows, the specific intensity goes to the Planck function $B_{\nu}$, and the Eddington tensor becomes
\begin{align}
\tensor{\pi}{^{(i)}_{(j)}} = \frac{\tensor{\delta}{^{(i)}_{(j)}}}{3}
\end{align}
When optical depths are not very large, $\tensor{\pi}{^{(i)}_{(j)}}$ is in general subject to few {\it a priori} constraints, and one must be able to evaluate Equation \ref{eqn:eddtensor}; that is, one must have knowledge of $I_{\nu}$.

\subsubsection{Fluid-Radiation Interactions}
\label{sec:fluidradinteractions}

Microphysical processes (emission, absorption, and scattering) lead to an exchange of four-momentum between the fluid and the radiation, along with a change in specific intensity along characteristics. 

The fluid and radiation stress energy tensors communicate through exchange of four-momentum. We can rewrite the divergence of the radiation stress-energy tensor as a four-force density,
\begin{align}
G_{\nu} \equiv -\tensor{R}{^{\mu}_{\nu;\mu}}.
\end{align}
From Equation \ref{eqn:gasrad}, we then have the interaction source terms on our fluid and radiation four-momenta:
\begin{align}
\left( \sqrt{-g} \tensor{T}{^{t}_{\nu}}\right)_{,t} &= \sqrt{-g} G_{\nu}
\label{eqn:radmomforce0}\\
\left( \sqrt{-g} \tensor{R}{^{t}_{\nu}}\right)_{,t} &= -\sqrt{-g} G_{\nu}.
\label{eqn:radmomforce}
\end{align}

In the fluid frame, the four-force density is (\citealt{MihalasMihalas1984})
\begin{align}
G_{(\mu)} = \int d\nu d\Omega \left( \alpha_{\nu}^{\rm a} I_{\nu} + \alpha_{\nu}^{\rm s} I_{\nu} - j_{\nu}^{\rm a}- j_{\nu}^{\rm s} \right) n_{(\mu)}.
\end{align}
We now rewrite this equation to produce something more amenable to stable integration. In particular, we want to use the $I_{\nu}$ to average source terms, rather than compute four-forces along individual characteristics as in a Monte Carlo approach, to reduce noise in the fluid. We will also introduce an angle-averaged approach to scattering.

The term involving $j_{\nu}^{\rm a}$ in the integrand does not depend on the intensity; we may integrate this emissivity directly ($J^{\rm a}$). In addition, we can without approximation rewrite the absorption terms as frequency- and angle-averaged opacities multiplying the comoving radiation four-momentum,
\begin{align}
G_{(0)} &= -\overline{\alpha}^{\rm a} \tensor{R}{^{(0)}_{(0)}} -\overline{\alpha}^{\rm s} \tensor{R}{^{(0)}_{(0)}} - J^{\rm a} - \int d\nu d\Omega j_{\nu}^{\rm s}
\label{eqn:fourforcetemp} \\
G_{(i)} &= \overline{\alpha}^{\rm a} \tensor{R}{^{(0)}_{(i)}} +\overline{\alpha}^{\rm s} \tensor{R}{^{(0)}_{(i)}} - \int d\nu d\Omega j_{\nu}^{\rm s} n_{(i)}
\end{align}
where
\begin{align}
\overline{\alpha}^{\rm a} &\equiv \frac{1}{I}\int d\nu d\Omega \alpha_{\nu}^{\rm a} I_{\nu} \\
\overline{\alpha}^{\rm s} &\equiv \frac{1}{I}\int d\nu d\Omega \alpha_{\nu}^{\rm s} I_{\nu}
\end{align}
\begin{align}
I \equiv \int d\nu d\Omega I_\nu\,.
\end{align}

We now consider the term involving $j_{\nu}^{\rm s}$ in Equation \ref{eqn:fourforcetemp}. For any $(d\Omega, d\nu)$, $j_{\nu}^{\rm s}$ can in principle depend on every other $(d\Omega, d\nu)$; all characteristics can scatter into each other. This is numerically very expensive unless treated in a probabilistic manner (e.g.\ \citealt{Dolence+2009}, \citealt{Ryan+2015}). Here we wish to preserve our continuum approach. Therefore, we instead derive a specific scattering emissivity from the evolution of an angle-averaged intensity given by the Kompaneets equation.

We approximate scattering by considering the angle-integrated transfer equation in the comoving frame with only source terms due to scattering
\begin{align}
\frac{d \mathcal{I}_{\nu}}{d s} = \mathcal{J}_{\nu}^{\rm s} - \mathcal{A}^{\rm s}_{\nu} \mathcal{I}_{\nu}
\label{eqn:avgdtransfereqn}
\end{align}
where script letters ($\mathcal{I}$, $\mathcal{J}$, $\mathcal{A}$) indicate a solid angle integral average. Both $ {d \mathcal{I}_{\nu}}/{d s}$ and $\mathcal{J}_{\nu}^{\rm s}$ are unknown. In order to proceed, we now specialize to electron scattering and introduce the Kompaneets equation (e.g.\ \citealt{RybickiLightman1979}), 
\begin{align}
\frac{\partial {n}}{\partial \tau} = n_{\rm e} \sigma_{\rm T} c \left(\frac{k_{\rm B} T_{\rm e}}{m_{\rm e} c^2} \right)\frac{1}{x^2} \frac{\partial}{\partial x} \left[ x^4 \left( \frac{dn}{dx} + n + n^2\right)\right]
\end{align}
where $n_{\rm e}$ is the electron number density, $\sigma_{\rm T}$ is the Thomson cross section, $T_{\rm e}$ is the electron temperature, $m_{\rm e}$ is the electron rest mass, $x = h \nu / (k_{\rm B} T_{\rm e})$, $n = c^2 \mathcal{I}_{\nu}/(2 h \nu^3)$ is the photon occupation number, and $\tau$ is here the proper time in the fluid frame. The Kompaneets equation is an angle-integrated (i.e.\ consistent with our scattering approximation) expansion of the Compton scattering kernel in the dimensionless energy transferred to photons per scattering event, $h \Delta \nu / (k_{\rm B} T_{\rm e})$. This value is small for nonrelativistic electrons, $T_{\rm e} \lesssim 10^8~{\rm K}$, and when small, Compton scattering becomes a diffusive flux in momentum space. At higher temperatures, one must take an integrodifferential approach to Compton scattering (e.g.\  \citealt{Jones1968}, \citealt{AharonianAtoyan1981}, \citealt{CoppiBlandford1990}, \citealt{Dolence+2009}, \citealt{Suleimanov+2012}, \citealt{Ryan+2015}, \citealt{Narayan+2017}, \citealt{Kinch+2019}).

Up to a multiplicative constant $\mathcal{C}$, $\partial n/\partial t \sim d \mathcal{I}_{\nu}/ds$. We can therefore use the Kompaneets equation to evaluate the change in intensity, and solve Equation \ref{eqn:avgdtransfereqn} for the angle-averaged scattering emissivity:
\begin{align}
\mathcal{J}_{\nu}^{\rm s} = \mathcal{C} \frac{\partial n}{\partial t} + \mathcal{A}^{\rm s}_{\nu} \mathcal{I}_{\nu}
\label{eqn:scatteringemissivity}
\end{align}
We can then evaluate the remaining integral in Equation \ref{eqn:fourforcetemp} by integrating $\mathcal{J}_{\nu}^{\rm s}$ over frequency, yielding our final expression for the radiation four-force:
\begin{align}
G_{(0)} &= -\overline{\alpha}^{\rm a} \tensor{R}{^{(0)}_{(0)}} -\overline{\alpha}^{\rm s} \tensor{R}{^{(0)}_{(0)}} - J^{\rm a} - 4 \pi \mathcal{J}^{\rm s} \label{eqn:fourforce} \\
G_{(i)} &= \overline{\alpha}^{\rm a} \tensor{R}{^{(0)}_{(i)}}+ \overline{\alpha}^{\rm s} \tensor{R}{^{(0)}_{(i)}}.
\end{align}

One obvious consequence of our angle-integrated procedure is that the integral over $j_{\nu}^{\rm s}$ does not contribute to $G_{(i)}$, because $j_{\nu}^{\rm s}$ is isotropic in the fluid frame. The error associated with this procedure depends on the differential cross section; for an isotropic scattering process, this is exact. The approximation we make to scattering is separate from any treatment of emission and absorption; those terms are solved exactly. Our approximate approach to scattering is not a fundamental requirement of MOCMC. We only invoke this approximation here for computational expediency and because this treatment is likely sufficient for some applications of immediate interest.

Equation \ref{eqn:fourforce} describes the exchange of four-momentum between fluid and radiation, but it does not update specific intensities. These are calculated by solving the transfer equation along each characteristic in invariant form. For each $I_{\nu}$ we solve Equation \ref{eqn:transfer}, with $j_{\nu}^{\rm s}$ approximated by $\mathcal{J}_{\nu}^{\rm s}$.

\section{Numerical Method}
\label{sec:method}

We now describe our numerical implementation of the equations of radiation MHD. Again, we divide our discussion into fluid, radiation, and interaction subsections.

\subsection{GRMHD Evolution}
\label{sec:grmhd}

Our method for integrating the GRMHD equations is \harm{} (\citealt{Gammie+2003}), a flux-conservative shock capturing scheme. All gas and magnetic field variables are zone-centered. Second-order accuracy in time is achieved with a midpoint method. The primitive variables are
\begin{align}
\left( \rho ,u_{\rm g} ,\tilde{u}^1 ,\tilde{u}^2 ,\tilde{u}^3,B^1,B^2,B^3\right)
\end{align}
where $B^i$ is the magnetic field three-vector and $\tilde{u}^i$ is related to the spatial components of the four-velocity, but more amenable to variable inversion (\citealt{McKinneyGammie2004}). The conserved variables $U$ are
\begin{align}
 \sqrt{-g}\left( \rho u^0 ,\tensor{T}{^0_{0}}-\rho u^0 ,\tensor{T}{^0_{1}} ,\tensor{T}{^0_{2}}  ,\tensor{T}{^0_{3}}, B^1, B^2,B^3 \right)
\end{align}
and the fluxes $F^i$ are
\begin{align}
\begin{split}
 \sqrt{-g}( &\rho u^i ,\tensor{T}{^i_{0}}-\rho u^i ,\tensor{T}{^i_{1}} ,\tensor{T}{^i_{2}}  ,\tensor{T}{^i_{3}}, \\
 &b^1u^i-b^iu^1, b^2u^i-b^iu^2,b^3u^i-b^iu^3 )
 \end{split}
\end{align}
Note that we have subtracted the rest mass from the time component of the stress-energy; this allows for greater accuracy in the internal energy of the fluid in certain cases. 

We use monotonized central (second order) or WENO5 (fifth order; \citealt{Liu+1994}, \citealt{Tchekhovskoy+2007}) methods to reconstruct primitive variables at zone faces. These in turn are used to calculate left and right fluxes $F_{\rm L}$ and $F_{\rm R}$ and conserved variables $U_{\rm L}$ and $U_{\rm R}$. The local Lax-Friedrichs approximate Riemann solver is then used to compute intercell fluxes of conserved variables,
\begin{align}
\mathcal{F}^i_{\rm g} = \frac{F_{\rm L}^i + F_{\rm R}^i - c_{\rm top,g}(U_{\rm R} - U_{\rm L})}{2},
\end{align}
where $c_{\rm top,g}$ is the maximum fluid wavespeed in the coordinate frame.

The no-monopoles condition is enforced via flux-CT (\citealt{Toth2000}). This method is robust and preserves a numerical discretization of Equation \ref{eqn:divb} to machine precision, although approaches using upwinded electromotive forces can deliver superior performance on at least certain problems (\citealt{GardinerStone2008}, \citealt{White+2016}) and can be extended to grids that are not logically Cartesian (\citealt{Duffell2016}).

\subsection{Radiation Moment Evolution}
\label{sec:radmom}

The conserved radiation four-momentum $\sqrt{-g}\tensor{R}{^{0}_{\nu}}$ is discretized spatially at zone centers. The samples in each zone will subsequently close the evolution equations for $\sqrt{-g}\tensor{R}{^{0}_{\nu}}$ by providing $\tensor{R}{^{i}_{j}}$. In this section we assume that the complete stress tensor is known. 

Similarly to the fluid evolution, we reconstruct from zone centers to faces using monotonized central or WENO5, and then calculate fluxes using the local Lax-Friedrichs solver. However, our radiation moment procedure differs from our treatment of advection for MHD in two important ways.

First, we reconstruct conserved variables and fluxes directly, rather than primitive variables. While reconstructing primitive variables is sometimes advantageous for hydrodynamic methods (preventing, for example, negative pressures) we have not found our alternative, more convenient approach to behave pathologically in the problems we have considered. In addition, transforming vector/tensor quantities like radiation fluxes or the Eddington tensor to locally orthonormal frames and then reconstructing them is not a unique process. While some choices of frames are obviously better than others from the standpoint of truncation error, free rotations in neighboring frames reconstructed to faces could introduce errors. Second, for the purposes of numerical diffusion we assume that the wavespeed is always $c$. This is accurate for free streaming along one direction, but for nearly isotropic radiation this will overestimate wavespeeds by a factor $\sqrt{3}$, leading to some additional numerical diffusion. 

Our fluxes are then given by (again, using local Lax-Friedrichs):
\begin{align}
\label{eqn:radfluxsplit}
{\mathcal F}^i_{\rm r} = \sqrt{-g}\frac{\tensor{R}{^{i}_{\nu}_{\rm ,L}} - \tensor{R}{^{i}_{\nu}_{\rm ,R}} + c \left( \tensor{R}{^{0}_{\nu}_{\rm ,R}}- \tensor{R}{^{0}_{\nu}_{\rm ,L}}\right)}{2}.
\end{align}

This approach can produce unacceptably large numerical diffusion when the optical depth per zone is very large: the diffusion applied by local Lax-Friedrichs overestimates that required for stability in what is essentially a parabolic problem, and Equation \ref{eqn:radfluxsplit} fails to recover the asymptotic diffusion limit. Previous authors (\citealt{JinLevermore1996}, \citealt{Sadowski+2013}, \citealt{Foucart+2015}, \citealt{Skinner+2019}) have treated this issue by calculating a separate flux valid in the optically thick limit, ${\mathcal F}^i_{\rm r,diff}$, and interpolating between these two based on the local optical depth. Here, we follow the procedure described in \cite{Foucart+2015}. Essentially, we split the energy flux into terms due to radiation advection and radiation diffusion, upwind the advection term, and average the diffusion term to construct ${\mathcal F}^i_{\rm r,diff}$, which we then smoothly interpolate towards from ${\mathcal F}^i_{\rm r}$ based on the intensity-weighted frequency-averaged optical depth in the zone $\Delta \tau_{\rm zone}$.

The lab frame energy flux in the Eddington closure, ${\mathcal F}^i_{\rm r,diff} = \tensor{R}{^i_0}$, with $u_{\rm r} \equiv \tensor{R}{^{(0)}_{(0)}}$, is (\citealt{Farris+2008})
\begin{align}
\tensor{R}{^i_0} = \frac{4}{3}u_{\rm r}u^iu_0 + F^iu_0 + u^i F_0
\end{align}
where $F^{\mu}$ is the radiation flux vector, defined such that $F^{\mu}u_{\mu} = 0$. We treat the first term as the advection of radiation energy by the fluid, and the subsequent terms as the diffusion of radiation. We adopt the relativistic expression for the flux (but drop time derivatives and the four-acceleration term),
\begin{align}
F_{i} = -\frac{1}{3 \chi \rho}\left(\tensor{\delta}{_i^j} + u_iu^j \right)u_{{\rm r},{j}}
\end{align}
and evaluate $F_{0}$ from the condition $F_{\mu}u^{\mu} = 0$. We then construct a final, stable, asymptotic-preserving flux ${\mathcal F}^i_{\rm r,asym}$ by interpolating based on the optical depth in the zone,
\begin{align}
{\mathcal F}^i_{\rm r,asym} = \epsilon {\mathcal F}^i_{\rm r} + (1 - \epsilon){\mathcal F}^i_{\rm r,diff}
\end{align}
where
\begin{align}
\epsilon = \tanh \left( \frac{1}{\Delta \tau_{\rm zone}} \right).
\end{align}

For the advective term in ${\mathcal F}^i_{\rm r,diff}$, we reconstruct the fluid primitive variables and $u_{\rm r}$ to each face. We use the reconstructed fluid primitive variables to evaluate fluid coordinate velocities. The average of the left and right coordniate velocities is used to define both the upwind direction and the advective velocity. The advection term is then evaluated using this average velocity and the reconstructed $u_{\rm r}$ on the upwind side of the face.

For the diffusive terms $F^i u_0 + u^i F_0$ in a flux ${\mathcal F}^j_{\rm r,diff}$, we first approximate $u_{{\rm r},i}$ along direction $j$ as 
\begin{align}
u_{{\rm r},i} \approx \frac{u_{\rm r}^{j} - u_{\rm r}^{j-1}}{\Delta x^i}.
\end{align}
We then evaluate the rest of the diffusive terms using left- and right-state fluid quantities, and take their average. 

\subsection{Samples and Geodesic Integration}
\label{sec:geodesic}

We discretize the invariant specific intensity $I_{\nu}/\nu^3$ with a swarm of samples. Each sample has a unique position $x^{\mu}$ and direction vector $n_{\mu}$, and carries an array of $I_{\nu}/\nu^3$ discretized logarithmically in $k_0$. In practice we initialize the positions and direction vectors by sampling uniformly in space and on the unit sphere in momentum space, respectively, but this is not a requirement. 

For stationary spacetimes (such as Minkowski space and rotating black holes) $k_0$ is invariant (and is equivalent to the frequency at infinity for asymptotically flat spacetimes). This would not hold in dynamical spacetimes, such as compact object mergers. In a tetrad frame with coordinate four-velocity $u^{\mu}$, the comoving frequency is $\propto u^{\mu}n_{\mu}$. Due to our logarithmic discretization in $k_0$, when considering frequency bins, we evaluate $\Delta \nu$ as $\nu \Delta \log \nu$, where $\Delta \log \nu$ is a constant. 

We evolve $x^{\mu}$ and $n_{\mu}$ by directly integrating Equation \ref{eqn:geodesic}, a set of ordinary differential equations, similarly to \cite{Dolence+2009}. We use the second-order-accurate Heun's method. We adaptively refine our integration to ensure some tolerance is met in $\Delta n^{\mu} n_{\mu}$ at each step; for the null geodesics we consider, $n^{\mu} n_{\mu}=0$, but this will not generally be conserved by our numerical geodesic integration. For spacetimes symmetric in $x^{\mu}$ the source terms on $n_{\mu}$ are zero; these quantities are conserved.

Source terms are applied to characteristics in an operator-split fashion after the geodesic update to $x^{\mu}$ and $n_{\mu}$ just described; see Section \ref{sec:interact}.

\subsection{Frame Transformations}
\label{sec:frames}

We employ two frames in this work: the coordinate frame, and a set of orthonormal tetrad frames comoving with the fluid. Essentially, the transport operators are evaluated in the coordinate frame, and source terms and the Eddington tensor are evaluated in comoving frames. These tetrads are constructed with Gram-Schmidt orthogonalization (e.g.\ \citealt{Dolence+2009}), producing transformation matrices between tetrad and coordinate frames
\begin{align}
\tensor{e}{^{(\mu)}_{\mu}} \\
\tensor{e}{_{(\mu)}^{\mu}}
\end{align}
accurate to roundoff error.

Fluid tetrads are constructed at the center of each zone. However, when boosting the samples inside a zone into that fluid frame, we construct a separate tetrad transformation at the spatial coordinate of that sample, and then simply interpret the axes of that tetrad as being ``close enough'' to the zone-centered fluid frame. This ensures that samples remain normalized in the zone-centered fluid frame; transforming a vector with a tetrad transformation evaluated at a different spatial coordinate will in general affect the normalization of that vector. However, this process is a source of error
in sample positions on the unit sphere in the comoving frame. Note that in curved space there is no unique way to assign angles between vectors that do not share spatial coordinates.

\subsection{Angular Reconstruction}
\label{sec:delaunay}

In the comoving frame, we wish to compute integrals of the specific intensity over solid angle. We treat the samples as support points and then define a quadrature rule.
A simple approach would be the traditional Monte Carlo method, in which every particle has equal angular ``weight'', and the sample intensities are simply summed over. Here, we adopt a more accurate approach, although MOCMC is largely agnostic in this respect.

\begin{figure}
\includegraphics[width=\columnwidth]{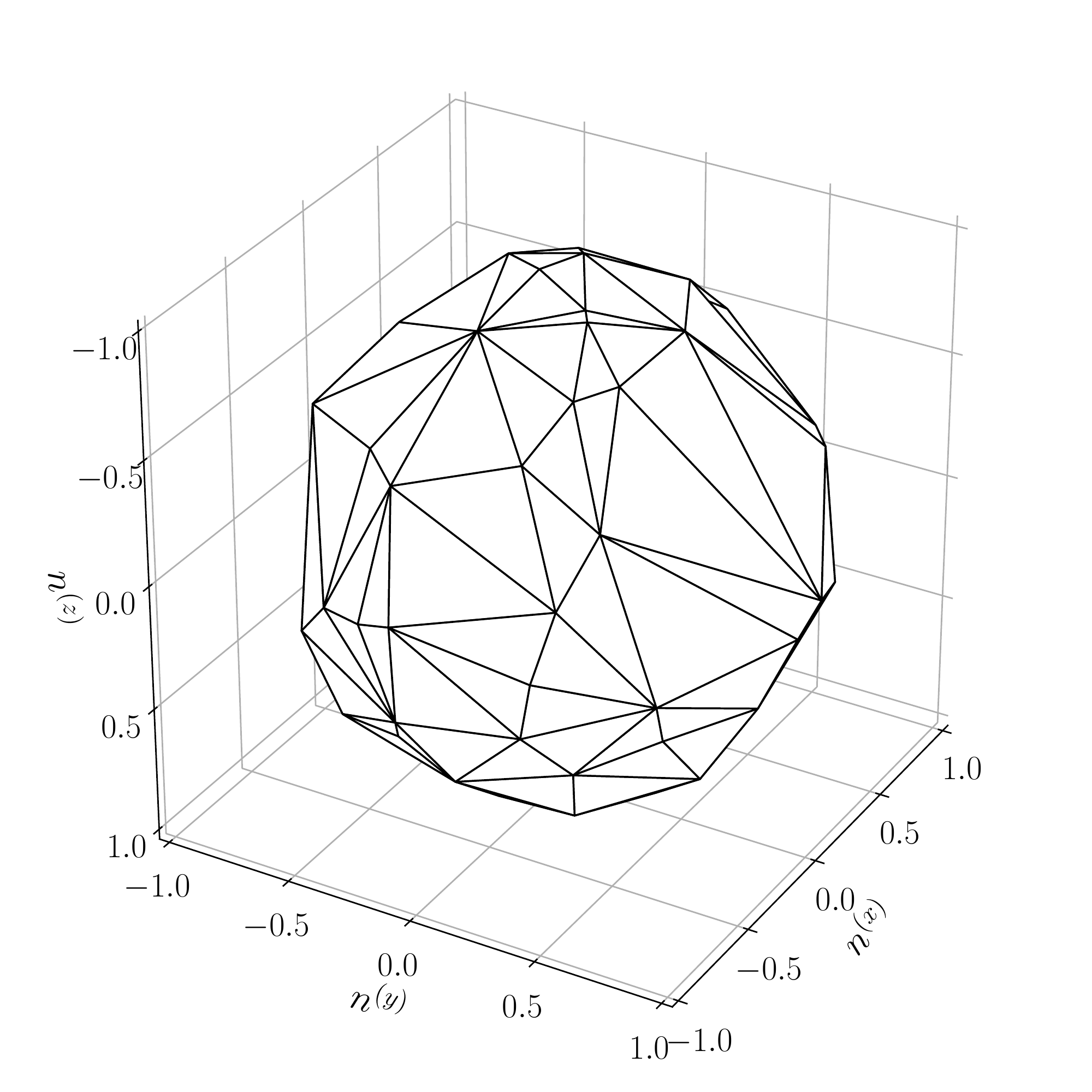}
\caption{The convex hull of 64 random points lying on the unit sphere. Vertices of triangles correspond to MOCMC samples. The projection of each triangle onto the unit sphere is the Delaunay triangulation of that sphere.}
\label{fig:hull}
\end{figure}

To reconstruct momentum space we construct a Delaunay triangulation, with the samples in each zone acting as vertices. To do this, we actually construct the convex hull, the smallest-volume region that contains the samples (or the closed surface one gets from ``shrinkwrapping'' the samples) in $\sim\mathcal{O}(N_{\rm samp} \log N_{\rm samp})$ time using the CGAL\footnote{CGAL, Computational Geometry Algorithms Library, \url{https://www.cgal.org}} library's implementation (\citealt{cgal}) of the quickhull algorithm (\citealt{Barber+1996}). $N_{\rm samp}$ is the number of samples in a zone. The facets of this hull, projected onto the unit sphere, are equivalent to the spherical Delaunay triangulation. Figure \ref{fig:hull} shows the convex hull of 64 points sampled uniformly on the unit sphere.

For each spherical triangle, we calculate the spherical excess $e$, or solid angle $\Delta \Omega$ subtended by the spherical triangle, with l'Huilier's theorem,
\begin{widetext}
\begin{align}
\Delta \Omega = 4 \tan^{-1} \left[\sqrt{\tan\left(\frac{s}{2}\right)\tan\left(\frac{s-\alpha}{2}\right)\tan\left(\frac{s-\beta}{2}\right)\tan\left(\frac{s-\gamma}{2}\right)}\right]
\end{align}
\end{widetext}
where $s = \left( \alpha + \beta + \gamma \right)/2$ and $\alpha$, $\beta$, $\gamma$ are the angles between each pair of vertices measured from the origin. We will use these $\Delta \Omega$ both to calculate the Eddington tensor and to angle-average intensities and opacities when evaluating fluid-radiation interactions.

For computational efficiency when evaluating angular quadratures over this triangulation, we interpolate the comoving-frame specific intensities for each sample onto a common frequency grid.

In the event that a zone contains fewer than four samples, the convex hull is not defined; instead, we set $\tensor{\pi}{^{(i)}_{(j)}} = \tensor{\delta}{^{(i)}_{(j)}}/3$ and $\Delta \Omega = 4 \pi/N_{\rm samp}$. In practice this is rare; our adaptive approach to sampling (Section \ref{sec:dynamicsampling}) generally prevents this for even a modest ($\sim 16$) average number of samples per zone.

\subsection{Eddington Tensor}
\label{sec:eddtensor}

The evaluation of $\tensor{R}{^{i}_{j}}$ is divided into two parts. First, we calculate the comoving frame $\tensor{\pi}{^{(i)}_{(j)}}$ by integrating the frequency-integrated intensities at the vertices of the triangulation over the unit sphere. Second, we calculate the comoving frame four-momentum such that when we transform the comoving frame radiation stress-energy tensor back to the coordinate frame, we recover the coordinate frame four-momentum we started with. Note that the evaluation of $\tensor{R}{^{(0)}_{(\mu)}}$ from $\tensor{R}{^{0}_{\mu}}$ depends on $\tensor{\pi}{^{(i)}_{(j)}}$, hence the complexity of this second part.

For integrating the Eddington tensor, we adopt a simple quadrature rule. For a spherical triangle with solid angle $\Delta \Omega$ and vertices with weights (here, frequency-integrated intensities) $w^1$, $w^2$, and $w^3$, the contribution to a component of the Eddington tensor is
\begin{align}
\Delta \tensor{\pi}{^{(i)}_{(j)}} = \Delta\Omega \frac{\sum\limits_m^3 \frac{w_m}{3}n^{(i)}n_{(j)}}{\sum\limits_m^3 \frac{w_m}{3}}
\end{align}
where the $n^{(i)}$, $n_{(j)}$ are evaluated at each vertex in the sum.

To increase accuracy at modest additional cost, we adopt the method of \cite{BoalSayas2004}, albeit with a different quadrature rule for individual triangles. For each initial triangular face, one may subdivide this triangle into four sub-faces and integrate these sub-faces individually. The resulting integral of a quantity over the sphere at a refinement level $N$ is then denoted $I_{N}$. \cite{BoalSayas2004} then conjecture that these refinement levels may be combined in a Richardson extrapolation-like manner to produce higher-order results, at least for weights known exactly (we simply average our weights to midpoints, although the $n^{(i)}$, $n_{(j)}$ we use are exact). We use one refinement level, i.e. the coarsest realization of this approach, which is conjectured to behave as:
\begin{align}
\int_{\Omega} = \frac{4}{3}I_{N+1} - \frac{1}{3}I_N + \mathcal{O}\left( \langle \Delta \Omega \rangle^4\right),
\end{align}
which is consistent with our limited numerical experiments. On nearly isotropic multi-zone problems in MOCMC, we have observed reductions in error in gas temperature of up to $30\times$ compared to results based on the single level integration scheme.

Given a set of triangles, the Eddington tensor is then 
\begin{align}
\tensor{\pi}{^{(i)}_{(j)}} = \frac{\sum\limits_l \Delta\Omega_l \sum\limits_m^3 \frac{I_m}{3} n^{(i)} n_{(j)}}{\sum\limits_l \Delta \Omega_l \sum\limits_m^3  \frac{I_m}{3}}
\end{align}
where $l$ indexes the triangulation faces, $m$ indexes the vertices of each triangle, $I_m$ is the frequency-integrated intensity at the $m$th vertex, and $n^{(i)}$ is the unit vector for each vertex.

We now have $R^0_{~\nu}$ in the coordinate frame and $\tensor{\pi}{^{(i)}_{(j)}}$ in the comoving frame; we want to recover a complete radiation stress tensor in the coordinate frame. The coordinate-frame four-momentum is related to the comoving frame stress-energy tensor by transformation matrices
\begin{align}
\tensor{R}{^0_{\nu}} = \tensor{e}{_{(\mu)}^0} \tensor{e}{^{(\nu)}_{\nu}} \tensor{R}{^{(\mu)}_{(\nu)}}.
\end{align}
$\tensor{R}{^{(i)}_{(j)}} = -\tensor{R}{^{(0)}_{(0)}}\tensor{\pi}{^{(i)}_{(j)}}$ and $\tensor{R}{^{(\nu)}_{(i)}} = -\tensor{R}{^{(i)}_{(\nu)}}$, so we can expand the right-hand side to recover a system of four linear equations for the unknowns $\tensor{R}{^{(0)}_{(\nu)}}$. By inverting a $4\times4$ matrix, we recover $\tensor{R}{^{(0)}_{(\nu)}}$, which in turn yields $\tensor{R}{^{(\mu)}_{(\nu)}}$ through $\tensor{\pi}{^{(i)}_{(j)}}$, which can then be transformed back to $\tensor{R}{^{\mu}_{\nu}}$.

\subsection{Dynamic sample resolution}
\label{sec:dynamicsampling}

An initial set of samples will stream out of outflow boundary conditions on the light crossing time of the simulation volume. Evidently, in such situations samples must be replenished. In addition, the number of samples per zone will fluctuate in multi-zone problems, especially when grids are irregular or gradients in Lorentz factor are appreciable, and we wish to ensure that the number of samples per zone never becomes too high or too low. Such a procedure also allows for dynamically controlling sample resolution: sensitive regions of integrations may require more samples than average.

Our approach to sample refinement and derefinement is simple and not unique. We first impose a desired number of samples per zone, which can in general depend on the local MHD or radiation properties. We then use this number of samples to define maximum and minimum solid angles for triangles on the unit sphere. 

If our triangulation results in a triangle with a solid angle above the maximum, we create a sample nearly at the center of this face and randomly positioned inside the zone with a specific intensity that is the average of the triangle's vertices. Samples are not created exactly at the face center to avoid providing the convex hull routine with two colocated vertices.

If our triangulation results in a triangle with a solid angle below the minimum, we identify the shortest edge of that triangle and randomly mark one of the two samples that form that edge for deletion. 

\subsection{Emission, Absorption, and Scattering}
\label{sec:interact}

Radiation interacts with plasma via emission, absorption, and scattering. For samples, these interactions are processed deterministically along characteristics. For the moment sector, we use sample intensities to frequency-average the opacities. Scattering is evaluated by using samples to construct an angle-averaged comoving spectrum, and evolving this spectrum with a scattering kernel. The change in four-momentum in the radiation moments determines the change in four-momentum of the fluid.

At large optical depths and/or small ratios of gas to radiation pressure, the characteristic timescale for four-momentum exchange between fluid and radiation can become much shorter than the global simulation timestep, and the problem becomes stiff. Strong coupling can be stabilized with implicit methods. For time-dependent radiation transport, implicit methods can be applied on a per-zone, or semi-implicit (i.e.\ the scheme remains explicit in spatial fluxes), basis, a major computational efficiency over global implicit solves. However, for intensities discretized in frequency and solid angle, such a semi-implicit solve could still require inverting a large matrix. 

Instead, we adopt the ``inner'' and ``outer'' loop approach of \cite{Skinner+2019}, in which one rootfinds over only one or a few nonlinear equations in an outer loop (indexed by $k$), and in each step of that rootfind updates intensities in a semi-implicit fashion using the most recent ($k^{\rm th}$) value for the gas temperature to evaluate emissivities and absorptivities.  We initialize this procedure by angle-averaging our sample intensities at timestep $n$, where we have used linear interpolation to shift the comoving sample intensities $I_{\nu}^n$ onto a common frequency grid in the comoving frame:
\begin{align}
\mathcal{I}_{\nu}^n = \frac{\sum\limits_l\left ( \Delta \Omega\right)_l I_{\nu}^{n}}{\sum\limits_l \left( \Delta \Omega\right)_l}.
\label{eqn:avgI}
\end{align}

\subsubsection{Inner Loop}

Our inner loop is composed of two steps. First, we use the Kompaneets equation to evaluate $\mathcal{J}_{\nu}^{{\rm s},k}$ (Equation \ref{eqn:scatteringemissivity}) and $\Delta u_{\rm r}^{{\rm s},k}$, the change in the comoving radiation energy density due to inelastic scattering, calculated as
\begin{align}
\Delta u_{\rm r}^{{\rm s},k} = \frac{4 \pi}{c}\mathcal{C}\sum_i \left( n_{i}^{n+1} - n_{i}^{n} \right) \nu_i \Delta \log \nu.
\end{align}
In general the integral of the $\mathcal{I}_{\nu}$ evaluated from the samples will not correspond to the moment sector's $u_{\rm r}$. Rather than normalizing $\mathcal{I}_{\nu}^n$ to recover $u_{\rm r}^n$, we normalize $\Delta u_{\rm r}^{\rm s} $ by the ratio of the energy density evaluated in the sample sector to that of the moment sector, avoiding issues with maintaining thermal spectra in the samples.

We adopt the numerical method of \cite{ChangCooper1970} for solving the Kompaneets equation. Here we briefly review this method. We discretize the equation for the evolution of photon occupation number $n$ over $x = h \nu / k_{\rm B} T_{\rm e}$ as
\begin{widetext}
\begin{align}
\frac{n_i^{n+1} - n_i^n}{\Delta \tau} &= n_{\rm e} \sigma_{\rm T} c \Theta_{\rm e} \frac{1}{x_i^3}\frac{1}{\Delta \log x_i} \left( x_{i+1/2}^4 F^{*}_{j+1/2} -  x_{i-1/2}^4 F^{*}_{j-1/2}\right)
\label{eqn:komp}
\end{align}
\end{widetext}
where $\Delta \tau$ is the elapsed proper time of the fluid from $n$ to $n+1$ and
\begin{align}
F^*_{i+1/2} &=\frac{n^{n+1}_{i+1} - n^{n+1}_{i} }{x_{i+1} - x_i}+ \left( 1 + n^n_{i+1/2}\right) n^{n+1}_{i+1/2} \\
n^n_{i+1/2} &= (1 - \delta_i)n^n_{i+1} + \delta_i n^n_i \\
n^{n+1}_{i+1/2} &= (1 - \delta_i)n^{n+1}_{i+1} + \delta_i n^{n+1}_i
\end{align}
where the $0 \leq \delta_i \leq 0.5$ are chosen to ensure stability using a quasiequilibrium distribution $n_{\rm eq}$,
\begin{align}
w_i &= \left(1 + n_{\rm{eq},j}/2 + n_{\rm{eq},j+1}/2\right)\left(x_{i+1} - x_i \right) \\
\delta_i &= \frac{1}{w_i} - \frac{1}{\exp \left(w_i\right) - 1}.
\end{align}
Note that $F^*_{i+1/2}$ contains both $n^{n+1}_i$ and $n^{n}_i$; this discretization is not fully implicit, a potential source of instability especially where $n$ is much greater than unity (for example, scattering of lines with brightness temperatures much greater than the fluid temperature). We enforce  $F^* = 0$ at the boundaries, thereby conserving photon number to machine precision. After each update we enforce a floor such that $n_i^{n+1} \geq 10^{-100}$. 

We solve this tridiagonal system of equations in $\mathcal{O}(N_{\rm bin})$ time, where $N_{\rm bin}$ is the number of frequency bins. This semi-implicit method can occasionally fail. We identify such situations by measuring the change in photon number density; if this quantity varies fractionally over a step by more than $10^{-8}$, we repeat the calculation without the nonlinear terms, yielding a fully implicit solution. One could also use an iterative method to solve the original equation in a fully implicit way; evidently the nonlinear terms are important when the semi-implicit method fails. However, we adopt our simpler approach to ensure stability and avoid nonlinear multidimensional rootfinding, at the potential cost of approaching Wien rather than Bose-Einstein photon distributions in scattering-dominated media. Other methods for discretizing the Kompaneets equation (e.g.\ \citealt{Larsen+1985}) may be less susceptible to this issue.

Another numerical difficulty is that calculation of the $\delta_i$ requires evaluation of a quasi-equilibrium solution. Here we follow \cite{ChangCooper1970} and choose a Bose-Einstein distribution
\begin{align}
n_{\rm eq} = \frac{1}{C e^x - 1}
\end{align}
where $C$, related to the photon chemical potential, is evaluated such that $n_{\rm eq}$ and $n$ yield the same photon number density. $n_{\rm eq}$ is singular at a point $x > 0$ for $C < 1$, which occurs when the photon number is greater than that of a blackbody at temperature $T_{\rm e}$. We avoid this difficulty by enforcing $C \geq 2$ (i.e.\ $n_{\rm eq} \leq 1$ everywhere) when calculating the $\delta_i$. We have not found this trick to damage stability or accuracy.

In the second step, we compute angle- and frequency-averaged opacities and integrated emissivities.  In the $k^{\rm th}$ iteration of the outer loop, we approximate the updated angle-averaged spectrum via a backward Euler discretization:
\begin{align}
\left( \frac{\mathcal{I}_{\nu}}{\nu^3}\right)^{k} = \frac{\frac{\mathcal{J}_{\nu}^{{\rm a},k} + \mathcal{J}_{\nu}^{{\rm s},k}}{\nu^2} + \frac{(\mathcal{I}_{\nu}/\nu^3)^n}{ \Delta s}}{\nu(\widetilde{\alpha}_{\nu}^{{\rm a},k} + \widetilde{\alpha}_{\nu}^{{\rm s},k})+ \frac{1}{\Delta s}}
\end{align}
where $\widetilde{\alpha}_{\nu}$ is the intensity-weighted, angle-averaged opacity at frequency $\nu$.  With these updated intensities in hand, we compute the frequency- and angle-averaged opacities:
\begin{align}
\overline{\alpha}^{{\rm a},k} &= \frac{\int d\Omega \sum \nu \Delta\log\nu I_\nu^{k} \alpha_\nu^{{\rm a},k}}{\int d\Omega \sum \nu \Delta\log\nu I_\nu^{k}} \\
\overline{\alpha}^{{\rm s},k} &= \frac{\int d\Omega \sum \nu \Delta\log\nu I_\nu^{k} \alpha_\nu^{{\rm s},k}}{\int d\Omega \sum \nu \Delta\log\nu I_\nu^{k}}
\end{align}
where the summation is over frequency bins. We also integrate the emissivity in a similar fashion, rather than analytically, so that truncation error in only absorption coefficient integration does not prevent the fluid from thermalizing effectively for finite numbers of frequency bins:
\begin{align}
J^{{\rm a},k} &= \int d\Omega \sum \nu\Delta\log\nu  j_{\nu}^{{\rm a},k}.
\end{align}
For clarity, we left these expressions in terms of integrals over solid angle.  In practice, we employ the same quadrature rule we described previously.

\subsubsection{Outer Loop}

In GRRMHD, the four-force update is four coupled equations. Previous authors have evaluated this in a fully nonlinear way with a 4D rootfind (\citealt{Roedig+2012}, \citealt{Sadowski+2013}, \citealt{McKinney+2014}) or a linearized 4D solve (\citealt{Foucart+2015}). Here, we adopt a more efficient approach, performing a 1D nonlinear solve for the gas energy density (e.g.\ \citealt{Skinner+2019}) in the comoving frame. We then construct an entire four-force $G_{(\mu)}$. 

For our 1D rootfind, we write the gas energy update implicitly, and solve iteratively for $u_{\rm g}^{n+1}$ with a secant method:
\begin{align}
{u_{\rm g}^{n+1} - u_{\rm g}^n}= -u_{\rm r}^{n+1}\left(\left(\frac{\mathcal{I}_{\nu}}{\nu^3}\right)^{n+1}, u_{\rm g}^{n+1} \right) + u_{\rm r}^n
\label{eq:implicitenergy}
\end{align}
where $u_{\rm r} = \tensor{R}{^{(0)}_{(0)}}$.  In the $k^{\rm th}$ iteration, we compute $u_{\rm r}^{k+1}$, our newest estimate of $u_{\rm r}^{n+1}$, from Equation \ref{eqn:fourforce} as
\begin{align}
u_{\rm r}^{k+1} = \frac{u_{\rm r}^n + \Delta \tau J^{{\rm a},k}}{1 + \Delta \tau \alpha^{{\rm a},k}} + \Delta u_{\rm r}^{{\rm s},k}.
\end{align}
Then we use Equation \ref{eq:implicitenergy} to evaluate a residual as
\begin{align}
r = (u_{\rm g}^k - u_{\rm g}^n) + (u_{\rm r}^{k+1} - u_{\rm r}^n)
\end{align}
and use the secant formula to compute our next guess for the gas energy, $u_{\rm g}^{k+1}$.  Convergence is reached when a sufficiently small residual is computed, at which point we set $u_{\rm g}^{n+1} = u_{\rm g}^{k+1}$.  This implicit approach maintains stability when energy exchange is significant over a timestep $\Delta \tau$. The secant method occasionally fails. In such cases, we repeat the rootfind using bisection. 

Once $u_{\rm g}^{n+1}$ is found, the radiation four-force is then calculated as 
\begin{align}
G_{(0)}^{n+1} &= - \frac{u_{\rm g}^{n+1} - u_{\rm g}^n}{\Delta \tau} \\
G_{(i)}^{n+1} &= \frac{\left(\left(\tensor{R}{^{(0)}_{(i)}}\right)^{n+1}  - \left(\tensor{R}{^{(0)}_{(i)}}\right)^{n} \right)}{\Delta \tau}
\end{align}
where $\tensor{R}{^{(0)}_{(i)}}$ is updated via backward Euler as
\begin{align}
\left(\tensor{R}{^{(0)}_{(i)}}\right)^{n+1} \equiv \frac{\left(\tensor{R}{^{(0)}_{(i)}}\right)^{n}}{1 + \Delta \tau \left( \overline{\alpha}^{{\rm a},n+1} + \overline{\alpha}^{{\rm s},n+1}\right)}\,.
\end{align}

The four-force is then transformed to the lab frame and applied to $T^0_{~\nu}$ and $R^0_{~\nu}$ using Equations \ref{eqn:radmomforce0} and \ref{eqn:radmomforce}. We also update the individual sample intensities in a backward Euler manner:
\begin{align}
\left( \frac{I_{\nu}}{\nu^3}\right)^{n+1} = \frac{j_{\nu}^{{\rm a},n+1}/\nu^2 + {\mathcal J}_{\nu}^{{\rm s},n+1}/\nu^2 + (I_{\nu}/\nu^3)^n/ \Delta s}{\nu\alpha_{\nu}^{{\rm a},n+1}  + \nu\alpha_{\nu}^{{\rm s},n+1}+ 1/\Delta s}.
\end{align}

From our experiments this approach shows excellent stability. Essentially, we exploit the different forms of $G_{(0)}$ and $G_{(i)}$ in the comoving frame using an operator split; while $G_{(0)}$ is the difference of an emission coefficient and an absorption coefficient, both of which can be large, three momentum exchange has only one term: $G_{(i)} = -\overline{\alpha} \tensor{R}{^{(0)}_{(i)}}$. Thus, while we rely on a numerical rootfind to evaluate the energy exchange, we can analytically evaluate the implicit momentum update. Since these interaction terms are typically treated with backward Euler for stability, this simple operator split does not degrade the temporal order of accuracy of the scheme.

\subsection{Time Integration}

Our method for time integration largely parallels the second order midpoint scheme in \harm{} (\citealt{Gammie+2003}). While we could process the radiation subsystem separately in a first-order fashion (e.g.\ \citealt{Jiang+2014a}, \citealt{Ryan+2015}, \citealt{Foucart2018}, i.e.\ one source term evaluation per timestep), we have found improved performance on transport tests when using a second-order-in-time update to the advective fluxes of the radiation stress-energy tensor. The midpoint method also allows larger CFL numbers than an Euler step, so it is unclear that we actually increase the cost of our scheme by going to second order accuracy in time for advective fluxes. However, source term updates are still first order in time. We typically use CFL = 0.8-0.9, independent of optical depth. Regardless of order of temporal accuracy, it is crucial, as we do here, to process in advance radiation interactions for any radiation moments used to source advective fluxes in order to correctly recover diffusion speeds in optically thick media.

Here we enumerate a complete timestep from $n$ to $n+1$. Tildes denote quantities which have been updated by radiation interactions at their current $n$.

\begin{enumerate}
\item Advect MHD conserved variables, $\widetilde{U}_{\rm g}^{n} \rightarrow U_{\rm g}^{n+1/2}$ sourced by $\widetilde{U}_{\rm g}^{n}$ (\S \ref{sec:grmhd}), and apply boundary conditions. 
\item Advect radiation moments,  $\left(\tensor{\widetilde{R}}{^{0}_{\nu}}\right)^n \rightarrow \left( \tensor{R}{^{0}_{\nu}}\right) ^{n+1/2}$ sourced by $\left(\tensor{\widetilde{R}}{^{\mu}_{\nu}}\right)^n$ (\S \ref{sec:radmom}), and apply boundary conditions. 
\item Calculate angle-averaged comoving intensity $\mathcal{I}^n_{\nu}$ (Equation \ref{eqn:avgI}), to be used for scattering in both Steps \ref{item:interact1} and \ref{item:interact2}.
\item Push samples along geodesics, $(x^i,n_{\mu})^n \rightarrow (x^i,n_{\mu})^{n+1/2}$ and $\left( \widetilde{I}_{\nu}/\nu^3\right)^n \rightarrow \left(I_{\nu}/\nu^3\right)^{n+1/2}$ (\S \ref{sec:geodesic}), and apply boundary conditions.
\item Boost samples to the fluid frame (\S \ref{sec:frames}) and construct triangulations in each zone (\S \ref{sec:delaunay}).
\item Integrate sample intensities over frequency and solid angle using triangles to evaluate  $\left(\tensor{\pi}{^{(i)}_{(j)}}\right)^{n+1/2}$  and then $\left(\tensor{R}{^{i}_{j}}\right)^{n+1/2}$ (\S \ref{sec:eddtensor}).
\item Calculate radiation four-force at $n+1/2$, apply to conserved MHD and radiation quantities, $U_{\rm g}^{n+1/2} \rightarrow \widetilde{U}_{\rm g}^{n+1/2}$ and $\left(\tensor{R}{^{0}_{\mu}}\right)^{n+1/2} \rightarrow \left( \tensor{\widetilde{R}}{^{0}_{\mu}}\right)^{n+1/2}$, and sample intensities $\left(I_{\nu}/\nu^3\right)^{n+1/2} \rightarrow \left(\widetilde{I}_{\nu}/\nu^3\right)^{n+1/2}$ (\S \ref{sec:interact}). \label{item:interact1}
\item Recalculate $\left(\tensor{\widetilde{\pi}}{^{(i)}_{(j)}}\right)^{n+1/2}$ with the $\widetilde{I}_{\nu}^{n+1/2}$ using the same triangulation, and use to evaluate $\left(\tensor{\widetilde{R}}{^{i}_{j}}\right) ^{n+1/2}$
\item Advect MHD conserved variables, $\widetilde{U}_{\rm g}^{n} \rightarrow U_{\rm g}^{n+1}$ sourced by $\widetilde{U}_{\rm g}^{n+1/2}$ (\S \ref{sec:grmhd}), and apply sources to sample boundary conditions. 
\item Advect radiation moments,  $\left( \tensor{\widetilde{R}}{^{0}_{\nu}}\right)^n \rightarrow \left( \tensor{R}{^{0}_{\nu}}\right)^{n+1}$ sourced by $\left(\tensor{\widetilde{R}}{^{\mu}_{\nu}}\right) ^{n+1/2}$ (\S \ref{sec:radmom}), and apply boundary conditions.
\item Push samples along geodesics, $(x^i,n_{\mu})^{n+1/2} \rightarrow (x^i,n_{\mu})^{n+1}$ and $\left(\widetilde{I}_{\nu}/\nu^3\right)^{n+1/2} \rightarrow \left(I_{\nu}/\nu^3\right)^{n}$ (\S \ref{sec:geodesic}), and apply boundary conditions.
\item Boost samples to the fluid frame (\S \ref{sec:frames}) and construct triangulations in each zone (\S \ref{sec:delaunay}).
\item Integrate sample intensities over frequency and solid angle using triangles to evaluate Eddington tensor $\left(\tensor{\pi}{^{(i)}_{(j)}}\right)^{n+1}$ and $\left(\tensor{R}{^{i}_{j}}\right)^{n+1}$  (\S \ref{sec:eddtensor}).
\item Calculate radiation four-force at $n+1$, apply to conserved MHD and radiation quantities, $U_{\rm g}^{n+1} \rightarrow \widetilde{U}_{\rm g}^{n+1}$ and $\left(\tensor{R}{^{0}_{\mu}}\right)^{n+1} \rightarrow \left(\tensor{\widetilde{R}}{^{0}_{\mu}}\right)^{n+1}$, and apply sources to sample intensities $\left(I_{\nu}/\nu^3\right)^{n+1} \rightarrow \left(\widetilde{I}_{\nu}/\nu^3\right)^{n+1}$ (\S \ref{sec:interact}). \label{item:interact2}
\item Recalculate $\left(\tensor{\widetilde{\pi}}{^{(i)}_{(j)}}\right)^{n+1}$ with the $\widetilde{I}_{\nu}^{n+1}$ using the same triangulation, and use to evaluate $\left(\tensor{\widetilde{R}}{^{i}_{j}}\right) ^{n+1}$
\end{enumerate}

\section{Tests}
\label{sec:tests}

We now consider a suite of tests including large and small optical depths, large and small ratios of radiation to gas pressures, relativistic motion, and curved spacetime. Apart from testing convergence, we focus on resolutions ($\sim 64$ samples per zone) that are realistic for global simulations.

For tests without periodic boundaries we adopt the resampling procedure described in Section \ref{sec:dynamicsampling} such that we add or remove samples in order to preserve approximately the same number of samples per zone for the duration of each simulation.

In several places we will compare MOCMC's performance on these tests to Eddington and M1 closures, and frequency-integrated (`gray') source terms. In doing so we focus on aspects in which discrepancies arise between angle- or frequency-averaged methods and transport solutions like MOCMC. Because MOCMC evolves the radiation four-momentum in order to conserve total four-momentum and the MOCMC samples act largely as a closure on the moment evolution equations, implementation of Eddington and M1 closures, along with gray opacities, into our method is straightforward. However, we restrict our moment implementation to emission, absorption, and isotropic elastic scattering; inelastic scattering (e.g.\ \citealt{Sadowski+2015}) is neglected. We use the Planck mean opacity for the source terms $G_{(\mu)}$, either with a gray opacity or the expression for bremsstrahlung emissivity given in \cite{RybickiLightman1979}.\footnote{As an example of an approach intermediate to a gray method like we adopt here and a frequency-dependent treatment,  \cite{Sadowski+2015} and \cite{Foucart+2016} also evolve a radiation particle number density, which can provide a characteristic radiation temperature.}

Briefly, we review here Eddington and M1 closures. These closures specify the spatial part $\tensor{R}{^{(i)}_{(j)}}$ of the radiation stress tensor. The Eddington closure assumes that the radiation is isotropic in the frame of the fluid; this is well-motivated at large optical depths because $I_{\nu} \rightarrow B_{\nu}$ as $\tau \rightarrow \infty$, and $B_{\nu}$ has no angular structure. This leads to 
\begin{align}
\tensor{R}{^{(i)}_{(j)}} = -\tensor{\delta}{^{(i)}_{(j)}}\tensor{R}{^{(0)}_{(0)}}/3.
\end{align}
The M1 closure, as often adopted, assumes that the radiation is isotropic in a frame not necessarily comoving with the fluid (\citealt{Levermore1984}, but see e.g.\ \citealt{Minerbo1978} for an alternative). The specific flux $f^i \equiv R^{0i}/R^{00}$ is used to calculate the frame of isotropy. In flat space this yields
\begin{align}
\label{eqn:M1mom}
\tensor{R}{^{(i)}_{(j)}} = -\left( \frac{1 - \xi}{2}\tensor{\delta}{^{(i)}_{(j)}} + \frac{3 \xi - 1}{2}\frac{f^{(i)} f_{(j)}}{f^{(k)} f_{(k)}}\right)\tensor{R}{^{(0)}_{(0)}}
\end{align}
where
\begin{align}
\label{eqn:M1xi}
\xi = \frac{3 + 4 f^{(k)} f_{(k)}}{5 + 2 \sqrt{4 - 3 f^{(k)} f_{(k)}}}.
\end{align}
For $f^{(i)} = 0$, this expression for $\tensor{R}{^{(i)}_{(j)}}$ recovers the Eddington closure. For pure streaming along a coordinate axis, e.g.\ $f^{(x)} = 1$, $\tensor{R}{^{(i)}_{(j)}} =- \tensor{\delta}{^{(i)}_{(x)}}\tensor{\delta}{^{(x)}_{(j)}}\tensor{R}{^{(0)}_{(0)}}$. The closure transitions smoothly between these limits. Note that this closure uses all the information available locally from the first moment (the conserved four-momentum, expressed here as $\tensor{R}{^{(0)}_{(0)}}$ and $f^{(i)}$). 

Our method for evaluating $\tensor{R}{^{i}_{j}}$ given $\tensor{R}{^0_{\mu}}$ and $\tensor{\pi}{^{(i)}_{(j)}}$ (Section \ref{sec:eddtensor}) does not generalize to the M1 closure, which depends on $\tensor{R}{^{(0)}_{(\mu)}}$. Instead, we use the approach of \cite{Sadowski+2013} to first recover $u_{\rm r}$ and the radiation four-velocity $u_{\rm r}^{\mu}$ from $\tensor{R}{^0_{\mu}}$. We then use their covariant expression for the radiation stress-energy tensor using the M1 closure,
\begin{align}
\tensor{R}{^{\mu}_{\nu}} = \frac{4}{3}u_{\rm r}u_{\rm r}^{\mu}u_{{\rm r},\nu} + \frac{1}{3}u_{\rm r} \tensor{\delta}{^{\mu}_{\nu}}
\end{align}
to evaluate $\tensor{R}{^{\mu}_{\nu}}$ in the coordinate frame. We then transform this quantity to the fluid frame with the tetrad transformations employed elsewhere in the code, where we recover  $\tensor{\pi}{^{(i)}_{(j)}} = \tensor{R}{^{(i)}_{(j)}}/\tensor{R}{^{(0)}_{(0)}}$.

When considering convergence, we disable adaptive sample refinement, to more finely control the sample resolution.

\subsection{Hohlraum Streaming}

\label{sec:hohlraum}

Consider a hohlraum boundary condition at $x = 0$ and temperature $T$ such that $I_{\nu} = B_{\nu}(T)$, and a vacuum for $x > 0$. The radiation energy density at the boundary is $u_{{\rm r},0} = a_{\rm r} T^4$. Radiation will propagate in the positive $x$ direction.

The time-dependent analytic solution is evaluated by sending characteristics backwards in time over all $\theta$ from position $x$ ($\theta = 0$ corresponds to the $+x$ direction) and determining whether they reach the hohlraum boundary by $t=0$. The specific intensity is then
\begin{align}
I_{\nu}   &= \begin{cases} 
      B_{\nu} & \theta < \theta_{\rm max} \\
      0 & {\rm else}
      \end{cases},
\end{align}
where $\theta_{\rm max} = \cos^{-1} \left( x/ct\right)$, the radiation energy density is
\begin{align}
u_{\rm r} = 2 \pi \int_0^{\infty} \int_0^{\theta_{\rm max}} B_{\nu} \sin \theta d \theta d\nu,
\end{align}
and the Eddington factor is
\begin{align}
\tensor{f}{^{(x)}_{(x)}} = 1/3\left[ \left( \frac{x}{ct}\right)^2 + \frac{x}{ct} + 1 \right].
\end{align}
At late time, $\theta_{\rm max} \rightarrow \pi/2$; while the intensity is far from thermal, the Eddington factor $\tensor{f}{^{(x)}_{(x)}} = 1/3$, exactly the value for $I_{\nu} = B_{\nu}$.
Although simple, this problem is analogous to physical situations encountered in radiation transport, in which a thermal object radiates into a tenuous atmosphere or ambient medium in nearly plane-parallel symmetry.

The hohlraum boundary condition in our simulation is enforced by setting the radiation moments to the value for a blackbody at temperature $T$. For MOCMC, we randomly distribute samples in the boundary zone at each timestep, with $I_{\nu} = B_{\nu}$ for these samples. However, because we are studying convergence on this problem, we want to avoid the discontinuity in radiation energy density at $x = 0$. Instead, we consider the boundary condition slightly away from $x=0$, where only characteristics moving in the $+x$ direction are thermal. Hence, our boundary condition is
\begin{align}
I_{\nu} &= \begin{cases} 
      B_{\nu} & \theta \leq \pi/2 \\
      0 & {\rm else} 
   \end{cases} \\
\tensor{R}{^{\mu}_{\nu}} &= \left[\begin{array}{cccc}
-u_{{\rm r},0}/2 & u_{{\rm r},0}/4 & 0 & 0	\\
-u_{{\rm r},0}/4 & u_{{\rm r},0} /6 & 0 & 0	\\
0 & 0 & u_{{\rm r},0}/6 & 0	\\
0 & 0 & 0 & u_{{\rm r},0}/6
\end{array}\right]
\end{align}
at $x = 0$. Note that Eddington and M1 produce similar results for this and an isotropic thermal boundary with $I_{\nu} = B_{\nu}$ and the corresponding $\tensor{R}{^{\mu}_{\nu}}$. The right boundary condition is placed at large enough $x$ to be causally disconnected from the simulation region shown. 

\begin{figure}
\includegraphics[width=\columnwidth]{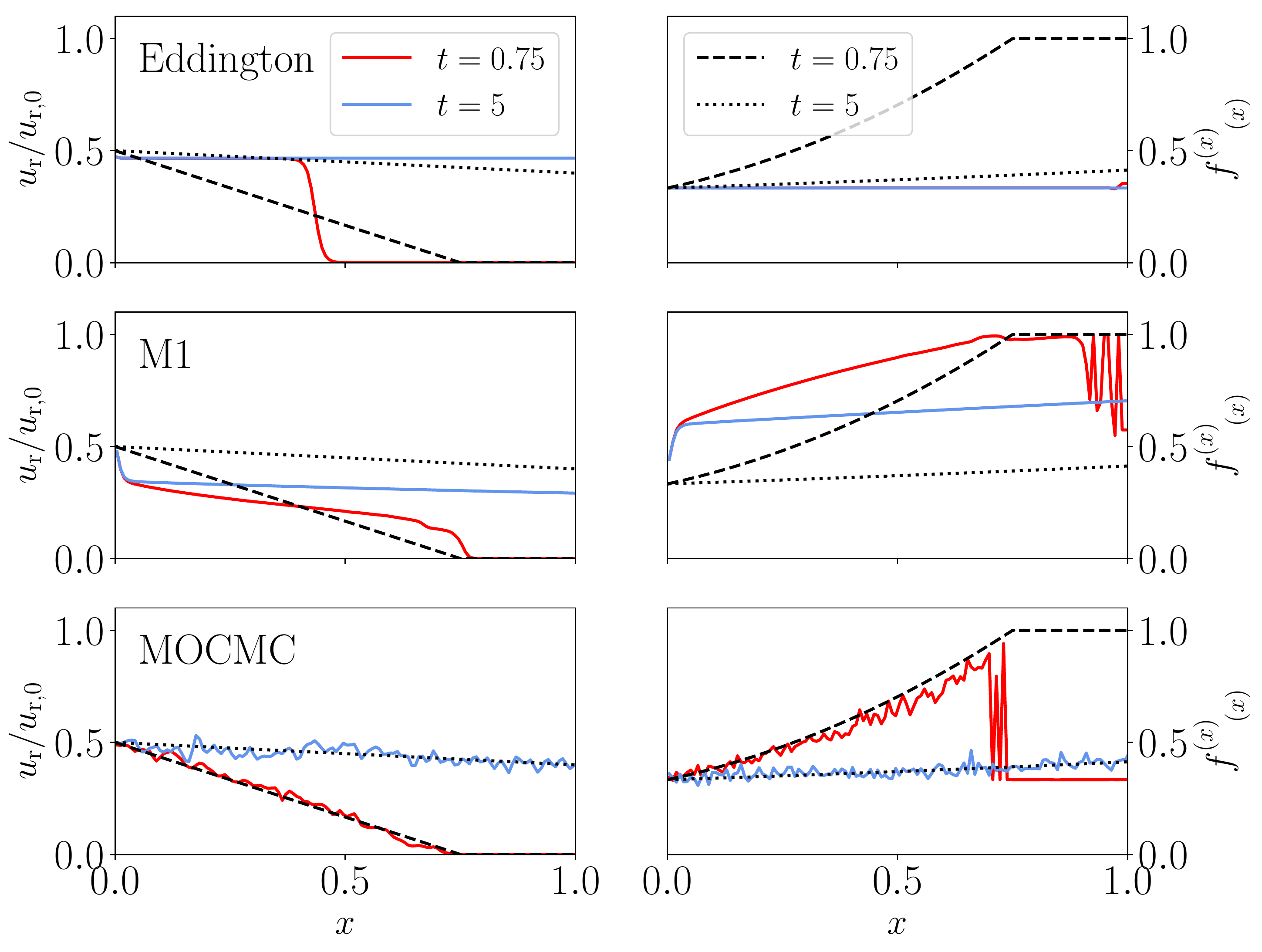}
\caption{Comparison for the hohlraum streaming test showing radiation energy density $u_{\rm r}$ and Eddington factor $\tensor{f}{^{(x)}_{(x)}}$ for Eddington, M1, and MOCMC methods at $t = 0.75$ for 128 zones between $x=0$ and $x=1$. The analytic solution at times $t = 0.75$ and $t=5$ is shown as black dashed and dotted lines; simulation results are shown as red and blue lines. The Eddington closure propagates as a pure wavefront moving at $v \sim c/\sqrt{3}$. 
M1 significantly overestimates $\tensor{f}{^{(x)}_{(x)}}$, because there is no frame in which the radiation is isotropic in this problem.
MOCMC (here with 60-80 samples per zone) agrees with the solution in both $u_{\rm r}$ and $\tensor{f}{^{(x)}_{(x)}}$, although it introduces noise. The bias in $\tensor{f}{^{(x)}_{(x)}}$ in the MOCMC solution is introduced by the interpolation of weights used when calculating $\tensor{\pi}{^{(i)}_{(j)}}$.}
\label{fig:hohlraum_compare}
\end{figure}

Figure \ref{fig:hohlraum_compare} compares Eddington, M1, and MOCMC closures on this test at $t=0.75$ and $t=5$. The assumption in M1, that there is a frame in which the radiation is isotropic, is not satisfied here. Additionally, $\tensor{f}{^{(x)}_{(x)}} = 1/3$ at late time while $f^{(x)} > 0$, which is inconsistent with Equations \ref{eqn:M1mom} and \ref{eqn:M1xi}.  Even in a time-independent sense, M1 leads to an order unity error in the radiation energy density, unlike the Eddington closure\footnote{In fact, at $t=\infty$ Eddington outperforms both M1 and MOCMC (by virtue of the lack of noise) on this problem.}. MOCMC accurately matches the analytic solution for both $u_{\rm r}$ and $\tensor{f}{^{(x)}_{(x)}}$. 

\begin{figure}
\includegraphics[width=\columnwidth]{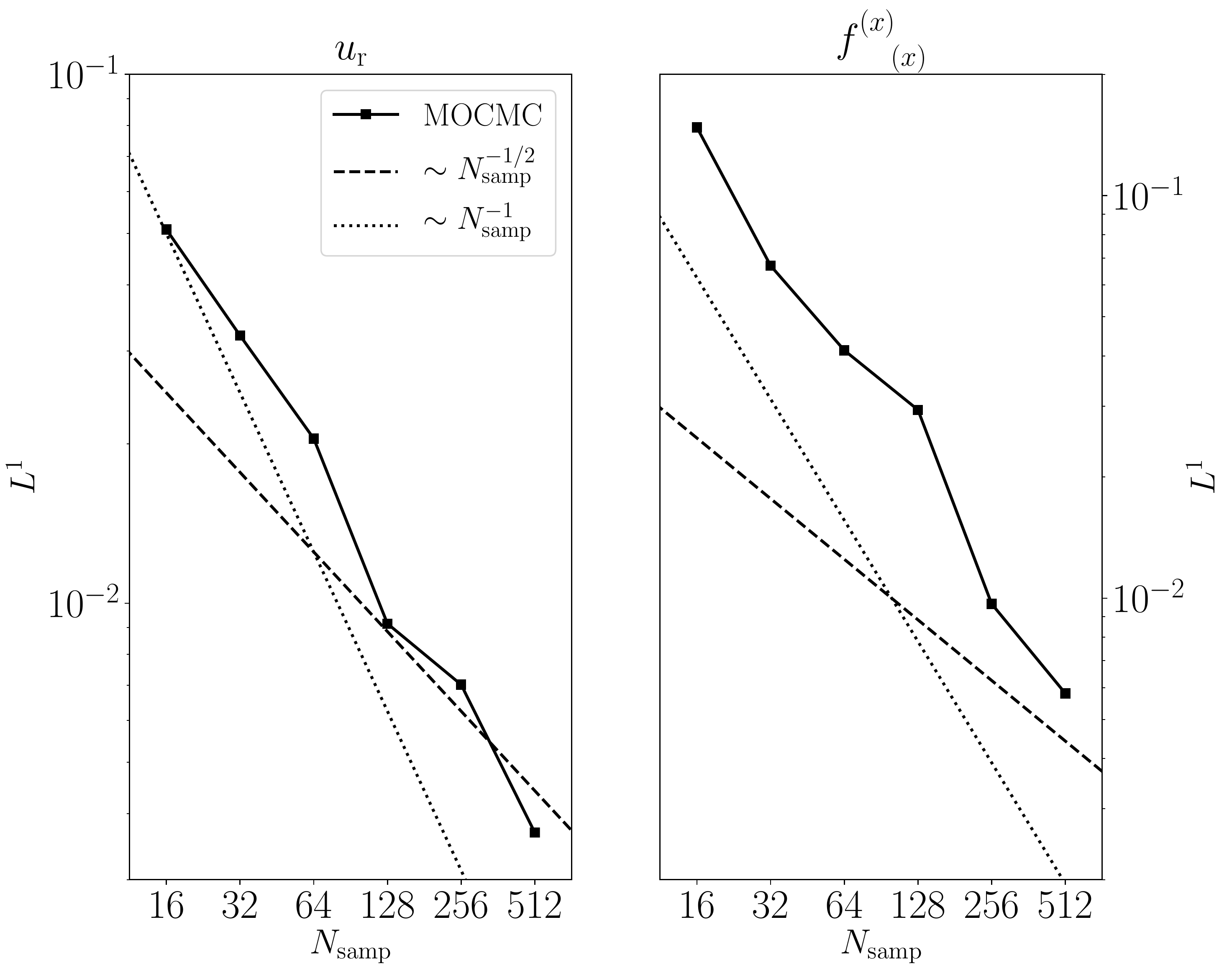}
\caption{Convergence of the hohlraum streaming test for MOCMC at $t=0.75$ with mean number of samples per zone $N_{\rm samp}$.  Convergence is nearly $N_{\rm samp}^{-1}$, rather than the Monte Carlo $N_{\rm samp}^{-1/2}$. Note that there is stochastic sampling error in this test; samples are initially distributed randomly on the unit sphere, and are randomly distributed at the thermal boundary.}
\label{fig:hohlraum_convergence}
\end{figure}

Convergence is shown in Figure \ref{fig:hohlraum_convergence}. Evidently we recover approximately first order convergence in $N_{\rm samp}$, unlike the $N_{\rm samp}^{-1/2}$ for a Monte Carlo method. This is a result of using the Delaunay triangulation to calculate $\tensor{\pi}{^{(i)}_{(j)}}$; essentially, the triangulation provides a more accurate estimate of the solid angle owned by each sample. We do not expect convergence better than first order on this test, which contains a discontinuity in momentum space. 

\subsection{2D Hohlraum}
\label{sec:2d_hohlraum}

Now consider a 2D box with hohlraum boundary conditions on both the $x=0$ and $y=0$ boundaries. 

We construct a solution for comparison via a simple Monte Carlo method. On a grid of positions $(x,y)$ and time $t$, we sample characteristics uniformly on the sphere, propagate them back to $t=0$, and ask whether they intersect a hohlraum boundary. Summing intensities over solid angle in each zone gives $u_{\rm r}$. 

\begin{figure}
\includegraphics[width=\columnwidth]{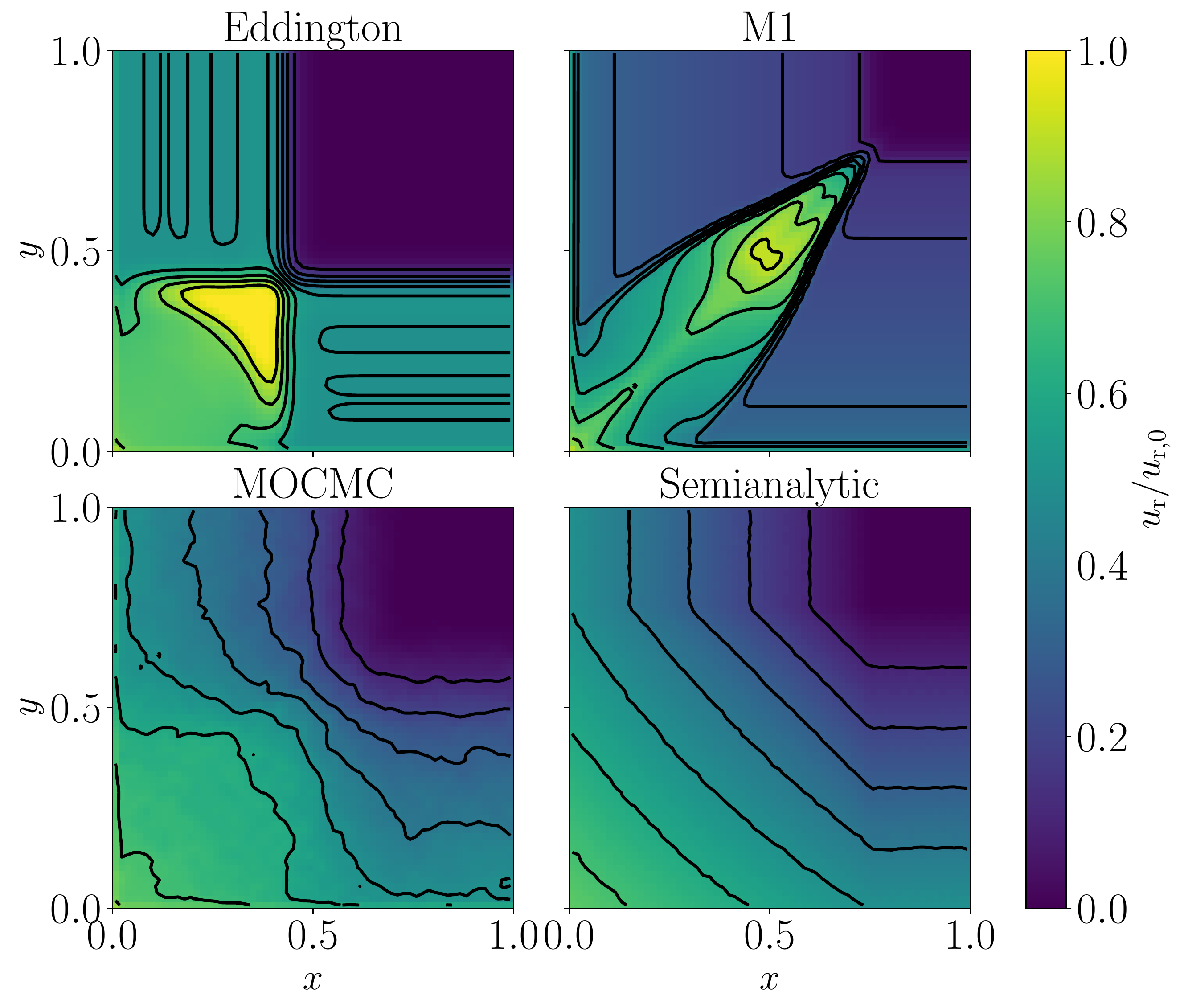}
\caption{Comparison for the 2D hohlraum streaming test with $64\times64$ zones showing radiation energy density at $t = 0.75$. Contours are spaced 0.1 apart. Moment closures differ qualitatively from the true solution. The Eddington closure generates a hot spot of radiation along the diagonal. M1 creates an even more dramatic jet of radiation along the diagonal. MOCMC recovers the semianalytic transport solution, although it introduces noise. The MOCMC solution used $\sim 64$ samples per zone. At finite time Eddington and M1 closures produce much stronger gradients in radiation energy density. Note that unlike a Newtonian diffusion equation, the Eddington closure to the radiation moments produces self-interaction like M1. At this resolution in samples, the MOCMC solution shows some radiation self-interaction as well due to truncation error; however, this decreases with increasing sample resolution.}
\label{fig:hohlraum_2d}
\end{figure}

Unlike in the 1D hohlraum test, the modified boundaries with nonzero fluxes are no longer valid, so we set the boundary conditions to simply be thermal at $x=0$ and $y=0$. The results for Eddington, M1, and MOCMC compared to the semianalytic solution are shown in \ref{fig:hohlraum_2d}.

This test shows that multidimensional effects are important to monitor when discriminating between transport algorithms. In particular, both moment methods exhibit significant interactions between the wavefronts from the two boundary conditions. These interactions lead to much larger radiation energy densities in the moment methods than are encountered in either MOCMC or the semianalytic solution. The MOCMC solution at 64 samples per zone also exhibits some radiation self-interaction due to truncation error in the Eddington tensor evaluation, but this self-interaction decreases with the number of samples and is already much lower than that seen in the moment methods.

\subsection{Thermalization}

\begin{figure}
\includegraphics[width=\columnwidth]{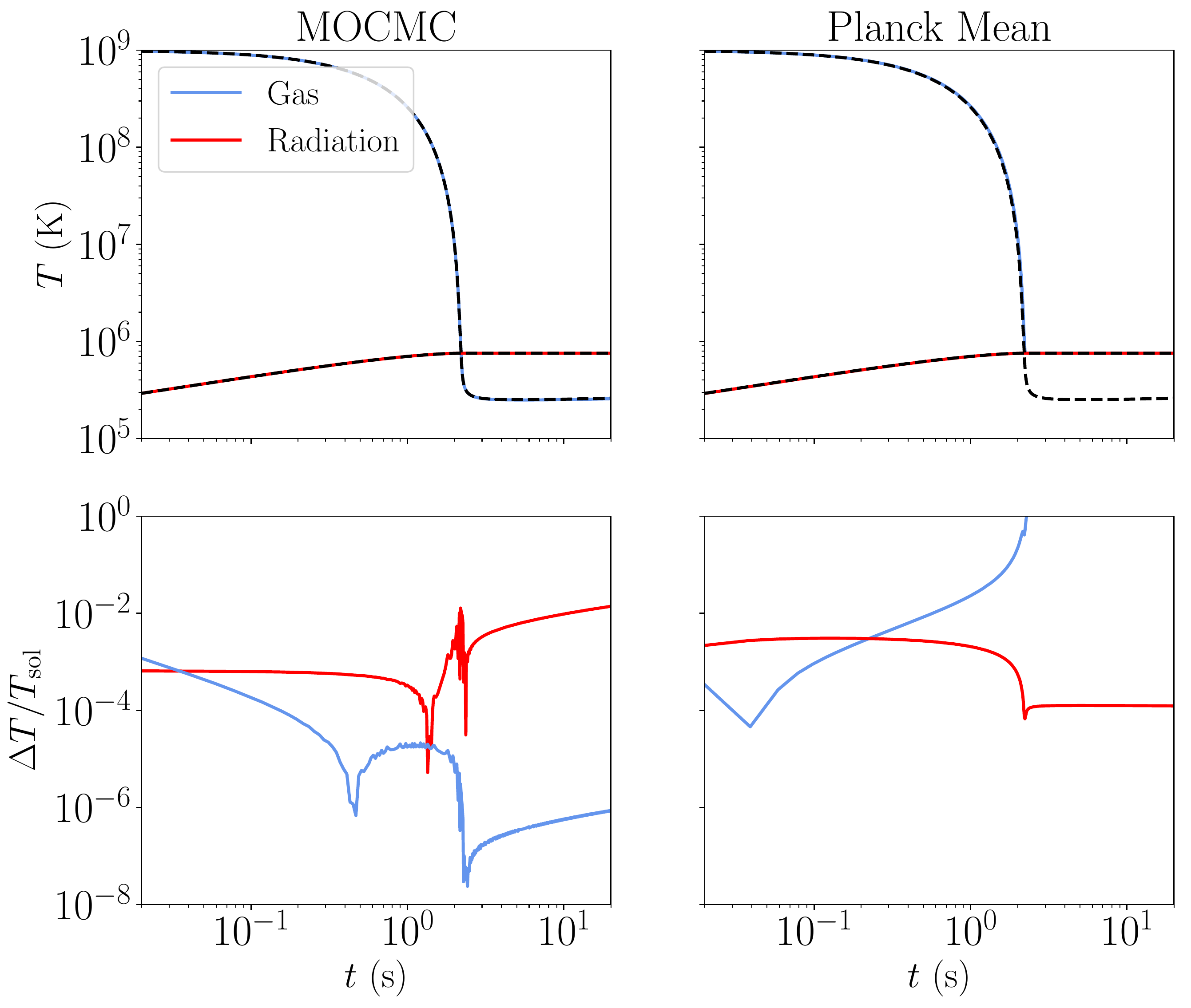}
\caption{Thermalization via bremsstrahlung, comparing a gray approach and the multigroup MOCMC method to the frequency-dependent semianalytic solution. The shifting $\exp \left( -h \nu / k_{\rm B} T\right)$ factor in the emissivity causes an undershoot in the gas temperature at intermediate times, as the initially emitted radiation is not easily reabsorbed by the now much colder gas. Gray methods that evolve only the radiation four-momentum, and so do not know about the frequency distribution, do not capture this effect, leading in this case to an order unity error in the gas temperature. Note that there is no error related to angular discretization in this isotropic problem. }
\label{fig:gray_multi_therm}
\end{figure}

Thermalization provides a useful test of the code's ability to recover a basic feature of radiation hydrodynamics, equilibration between material and radiation temperatures. We repeat the thermalization test from \cite{Ryan+2015} on a $3\times3$ grid of spatial zones, with a much larger initial gas temperature $T_{{\rm g,}0} = 10^9~{\rm K}$ and electron number density $n_{\rm e} = 6\times10^{16}~{\rm cm^{-3}}$. There is no radiation initially. The gas and radiation are allowed to proceed towards equilibrium. We compare to a frequency-dependent semianalytic solution in Figure \ref{fig:gray_multi_therm}.

We also compare to frequency-integrated source terms i.e.\ we use a Planck mean opacity rather than an opacity averaged over the samples. Because the exponential cutoff in the emissivity shifts dramatically downwards in frequency, the timescale to thermalize radiation that is emitted initially is very long, resulting in an undershoot of the gas temperature. This is not captured by a gray method, which cannot know the frequency distribution of radiation. Evidently this effect will be less important for systems that do not deviate far from equilibrium.

\subsection{Spiegel Linear Mode}

The time evolution of a temperature perturbation in an otherwise uniform medium in radiative equilibrium with gray absorption coefficient $\alpha$ permits an exact solution under the assumption that the time-dependent terms in the radiative transfer equation are small (\citealt{Spiegel1957}, \citealt{MihalasMihalas1984}). Note that this test assumes that light is transported much faster than the mode decay time; attempting to recreate this test in a slow-light code like ours will generally introduce some numerical diffusion through the Riemann solver in the radiation sector.

Unlike \cite{Ryan+2015}, here we hold $\alpha_0$, $T_0$, and $c t_{\rm RR} / L$ constant. This problem requires that the relaxation rate be slow compared to the light crossing time for the perturbation wavelength. We use 32 grid zones to simulate one wavelength.

\begin{figure}
\includegraphics[width=\columnwidth]{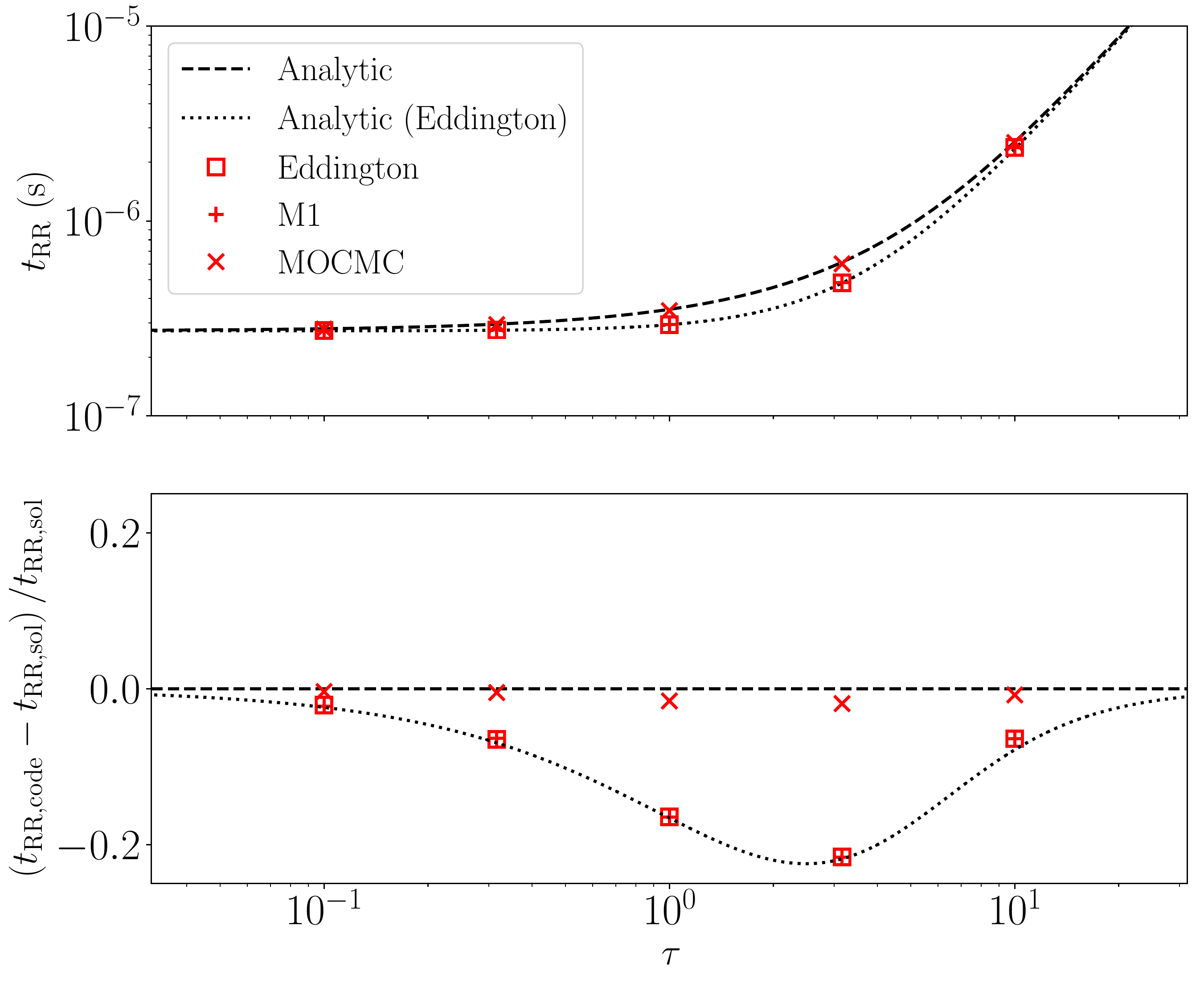}
\caption{Dispersion relation, and fractional error, as a function of optical depth per wavelength $\tau$ for the Spiegel linear mode test. Analytic solutions are shown, both for full transport and using the Eddington approximation. M1 and Eddington are essentially identical in this test; the anisotropy in the intensity is only perturbative so $f^{(x)} \ll 1$ always, and for the full transport solution, the Eddington factor $\tensor{f}{^{(x)}_{(x)}}$ of the perturbation varies between 0 and $1/3$, whereas for M1, $\tensor{f}{^{(x)}_{(x)}} \in [1/3, 1]$. MOCMC shows good agreement with the full transport solution.}
\label{fig:spiegel_dispersion}
\end{figure}

We measure the relaxation rate by fitting the form of the linear solution to the numerical result after one e-folding time. We compare to the gas temperature only, because the perturbed radiation energy density is not trivial when the optical depth is finite. We show the performance across optical depth by comparing measured dispersion rates for Eddington, M1, and MOCMC to the analytic solutions, both for transport and for the Eddington closure, in Figure \ref{fig:spiegel_dispersion}. 

We also show the analytic solution for the Eddington approximation. For the moderately optically thick limit, these closures produce similar errors, $\lesssim 20\%$. M1 does not distinguish between isotropic and anisotropic components, and the isotropic component dominates the four-momentum used to compute $\tensor{f}{^{(x)}_{(x)}}$.  

The Eddington factor for the perturbation is 
\begin{align}
\tensor{f}{^{(x)}_{(x)}}\left( \tau \right) = \frac{\tau}{2 \pi}\frac{1 - \frac{\tau}{2 \pi} \tan^{-1} \left( \frac{2 \pi}{\tau}\right)}{\tan^{-1} \left( \frac{2 \pi}{\tau}\right)};
\end{align}
for $\tau \rightarrow \infty$, $\tensor{f}{^{(x)}_{(x)}} \rightarrow 1/3$; as expected, the Eddington approximation is appropriate. For $\tau \rightarrow 0$, however, $\tensor{f}{^{(x)}_{(x)}} \rightarrow 0$, whereas Eddington and M1 closure both assume $\tensor{f}{^{(x)}_{(x)}} \geq 1/3$; the perturbation in the radiation field becomes oblate along the $x$ axis. Eddington factors less than $1/3$ also appear in radiative shocks (\citealt{Jiang+2014a}). Clearly standard moment closure models are invalid here, and yet such closures recover the correct dispersion relation at small $\tau$. \cite{MihalasMihalas1984} point out this is because, for $\tau \rightarrow 0$, relaxation is determined entirely by emission.

\subsection{Comptonization}

We repeat the setup from \cite{Ryan+2015} of thermalization of soft photons due to Compton upscattering in a one-zone box. For this test we use 32 samples per zone, although there is no angular structure, so there is no truncation error due to solid angle discretization (note the absence of noise in the solution). We use 100 frequency bins, logarithmically spaced from $10^8~{\rm Hz}$ to $10^{20}~{\rm Hz}$. We initialize the gas with electron and proton number densities $n_{\rm e} = 2.5\times10^{17}~{\rm cm^{-3}}$ and temperature $T_{{\rm g},0} = 5 \times 10^7 ~{\rm K}$. The radiation is initially monochromatic at frequency $\nu_0 = 3\times10^{16}~{\rm Hz}$ and photon number density $n_{\gamma} = 2.38\times10^{18}~{\rm cm^{-3}}$. 

\begin{figure}
\includegraphics[width=\columnwidth]{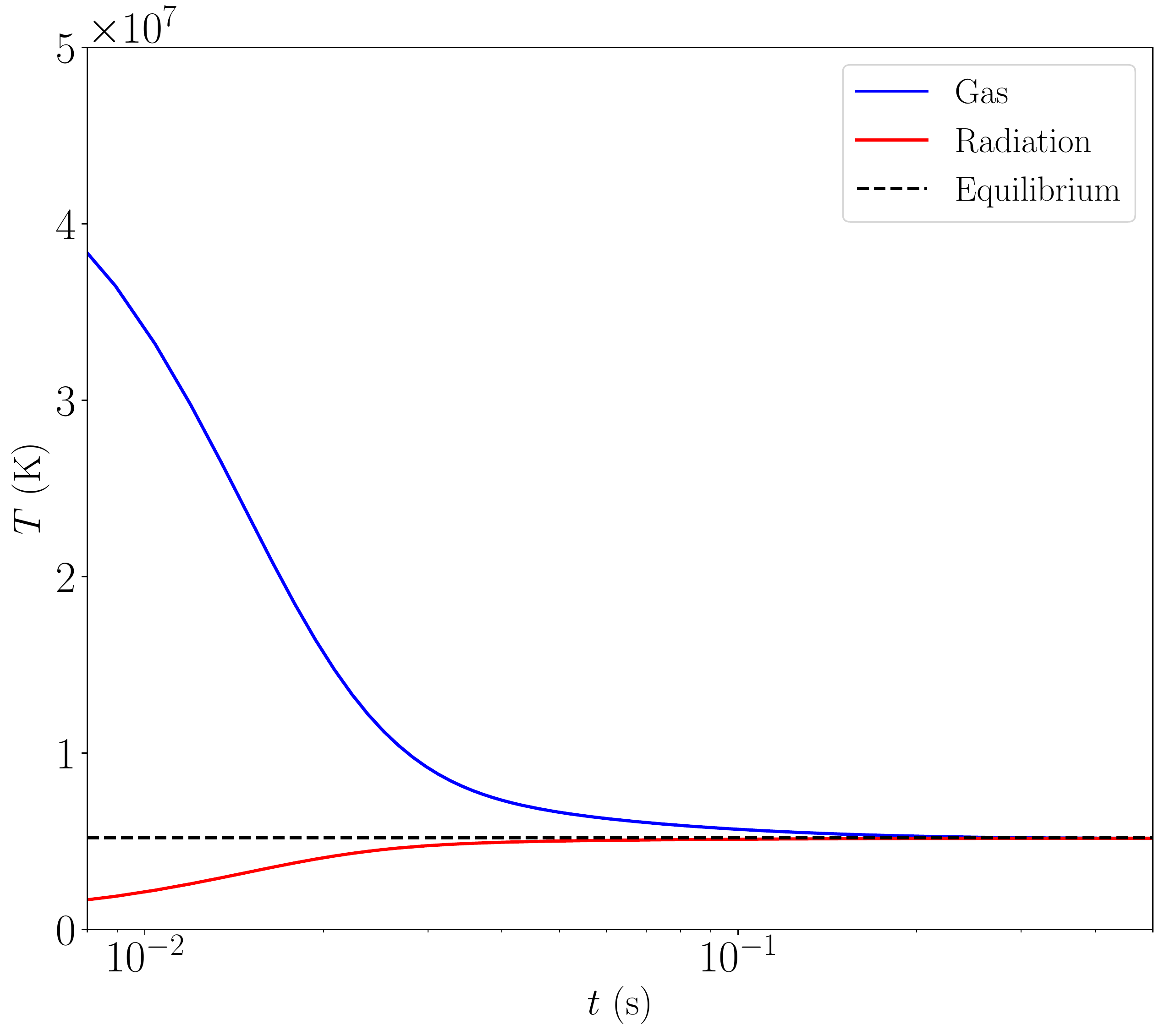}
\caption{Comptonization of soft monochromatic photons. The equilibrium temperature is shown as the black dashed line. MOCMC equilibrates to the correct temperature.}
\label{fig:comptonization}
\end{figure}

We calculate the equilibrium temperature with conservation of energy and photon number. Unlike \cite{Ryan+2015}, here we assume that the final photon distribution is Bose-Einstein rather than Wien; the final temperature $T_f$ is found by solving
\begin{equation}
\begin{aligned}
\frac{2 n_{\rm e} k_{\rm B} T_{{\rm g},0}}{\gamma - 1} + h \nu_0 n_{\gamma} = \\
\frac{2 n_{\rm e} k_{\rm B} T_{\rm f}}{\gamma - 1} + \frac{48 \pi k_{\rm B}^4 T_{\rm f}^4}{c^3 h^3}{\rm Li}_4 \left(  \exp \left(\frac{- \mu }{k_{\rm B} T_{\rm f}}\right)\right)
\end{aligned}
\end{equation}
where $\mu$ is the chemical potential of the photons and ${\rm Li}_s\left( z\right)$ is the polylogarithm. For this test, the photon occupation number remains much less than one, and $T_f$ closely agrees with the value calculated assuming a final Wien distribution for the photons. The result is shown in Figure \ref{fig:comptonization}.

\subsection{Static Diffusion}

\begin{figure}
\includegraphics[width=\columnwidth]{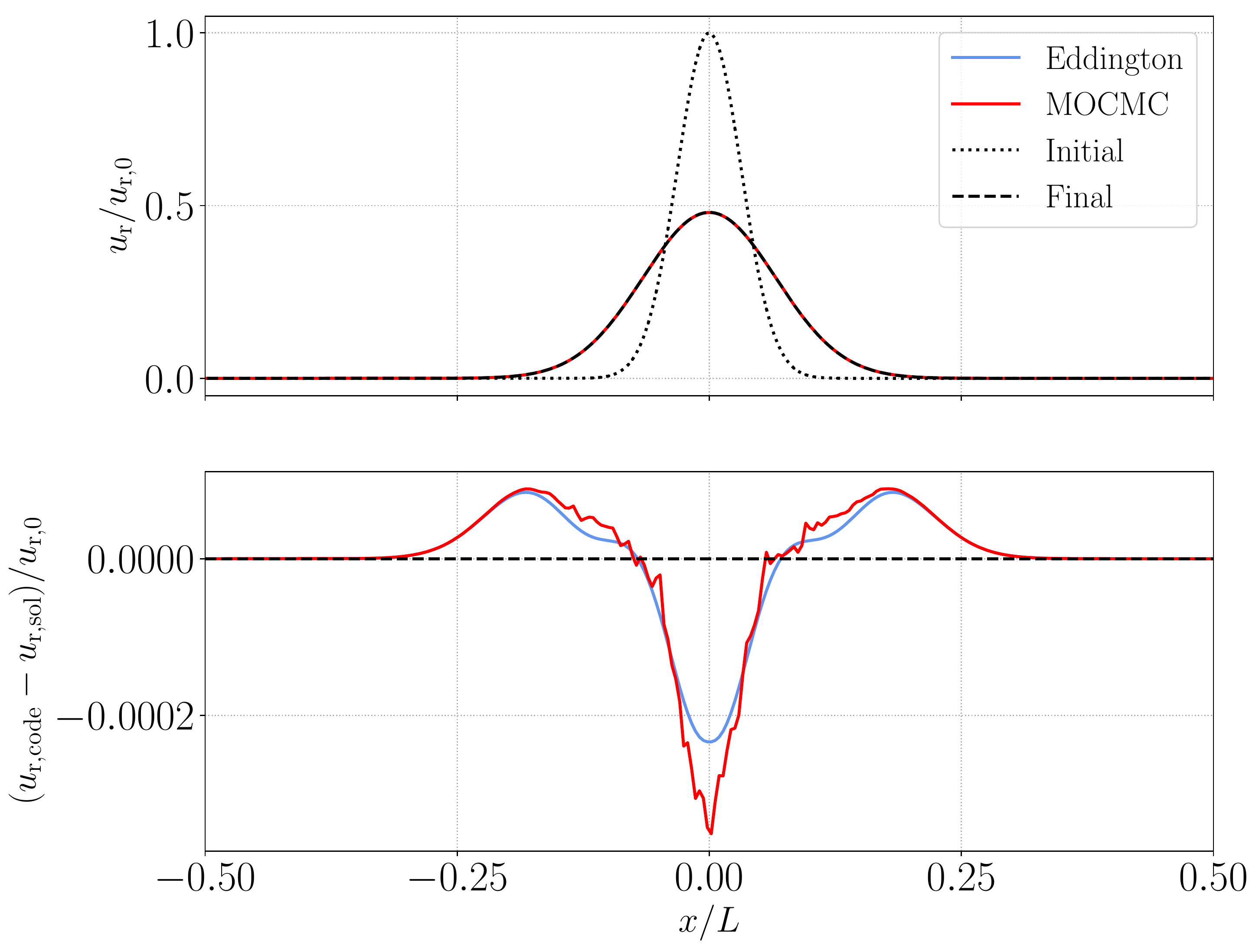}
\caption{Static diffusion of a Gaussian pulse in a uniform medium optically thick to Thomson scattering ($\tau = 10^4$) in MOCMC with 256 zones and 64 samples per zone. The top panel shows the initial and final analytic solutions relative to the maximum initial radiation energy density $u_{{\rm r},0}$, along with the MOCMC solution, and the bottom panel shows the residuals, at the few \% level where $u_{\rm r}/u_{{\rm r},0}$ is significant, in the radiation energy density at the final time.}
\label{fig:static_diffusion}
\end{figure}

To test the performance of our scattering treatment, and the behavior of MOCMC in the diffusion regime, we consider diffusion of a Gaussian pulse in a static medium optically thick to Thomson scattering in 1D on a domain $x \in \left[ -L/2, L/2\right]$. Starting at $t_0$, the analytic solution for the radiation energy density (with a 0.01\% background) in the diffusion regime is
\begin{align}
u_{\rm r} = u_{{\rm r},0} \left(10^{-4}+ \sqrt{\frac{t_0}{t}}\exp \left( \frac{-x^2}{4 D c t}\right) \right)
\end{align}
where the diffusion coefficient $D = 1/(3 \sigma_{\rm T} n_{\rm e})$ and $u_{{\rm r},0}$ is the maximum radiation energy density at the initial time $t=t_0$. We set $n_{\rm e}$ such that the optical depth over the domain is $\tau = 10^4$. We evolve this system to $t = 50 L/c$; the MOCMC solution is shown in Figure \ref{fig:static_diffusion}. The solution shows good agreement; in particular, the pulse diffuses much more slowly than the rate at which samples traverse zones.

\subsection{Dynamic Diffusion}

To study dynamic diffusion, we add a lab-frame velocity to the fluid such that the radiation is advected with the flow more rapidly than it diffuses. This is a powerful test of accuracy for a radiation hydrodynamics method. In particular, this test can be a challenge for $\mathcal{O}\left( v/c \right)$ methods when care is not taken in truncating the equations in a way appropriate for all regimes (\citealt{Krumholz+2007}). This is captured naturally by covariant methods like MOCMC.

\begin{figure}
\includegraphics[width=\columnwidth]{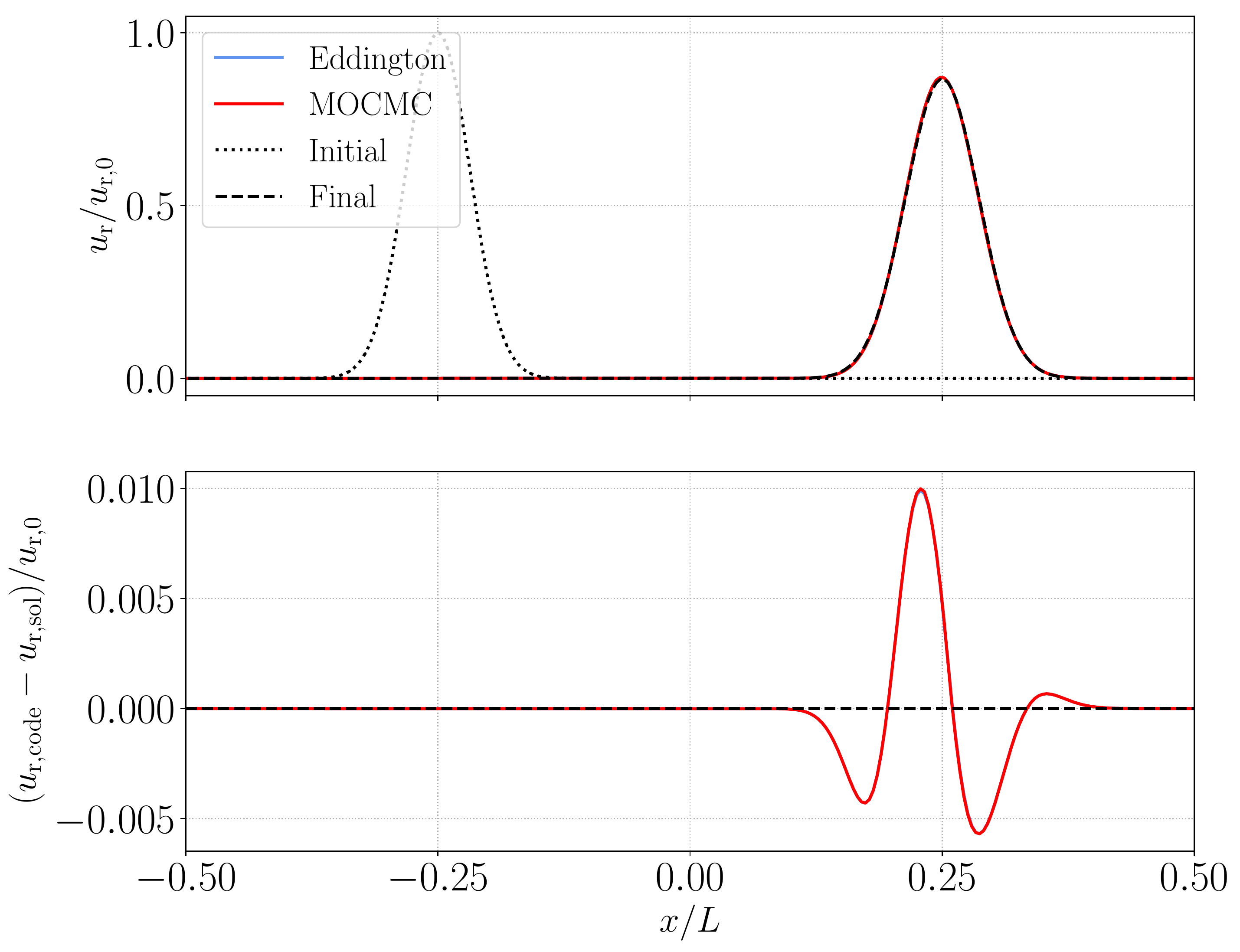}
\caption{Diffusing pulse due to Thomson scattering with MOCMC. The fluid is moving with speed $0.1c$. The top panel shows radiation energy density in red at the final time, along with the initial conditions (black dotted line) and analytic solution at the final time (black dashed line). The bottom panel shows the fractional error in the Eddington and MOCMC solutions. The optical depth across the domain is $10^4$. 256 zones were used, with 64 samples per zone. The particle noise in constructing the Eddington tensor is negligible compared to other errors in this problem; the Eddington and MOCMC solutions are nearly indistinguishable in this plot.}
\label{fig:dynamic_diffusion}
\end{figure}

We adopt the domain and initial conditions from the previous test, except that the pulse is now initially centered at $x/L = -0.25$, we set the fluid speed to $\beta \equiv v/c = 0.1$, and we evolve the system to $t _f = 5 L/c$. As in the previous test, we set $\tau = 10^4$; at $t_f$ there is some diffusion of the pulse. Figure \ref{fig:dynamic_diffusion} shows the MOCMC solution, along with initial and final analytic solutions.

\subsection{Noisy Equilibrium}

We consider a 1D box of length $L$ with gas and radiation initially in thermal equilibrium at temperature $T_0 = 10^7~{\rm K}$. As this system evolves, noise in the Eddington tensor will lead to noise in the radiation four-momentum, which in turn will couple to the gas energy density. We use this test to measure the noise in the fluid when varying the optical depth across the box $\tau$ and the gas to radiation pressure ratio $\beta_{\rm r}$. Note that noise-free methods like analytic moment closures satisfy this test trivially.

\begin{figure}
\includegraphics[width=\columnwidth]{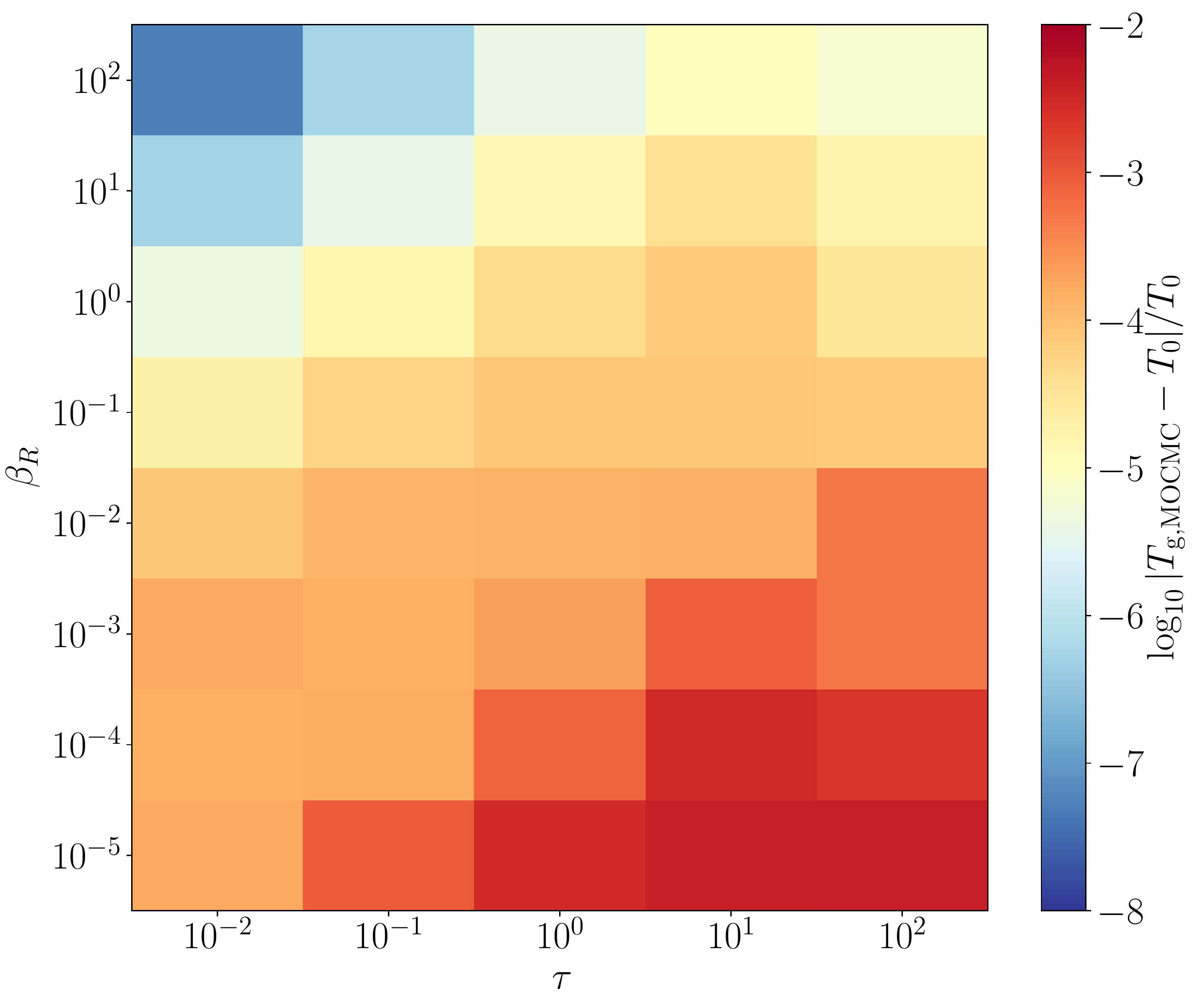}
\caption{Average relative error in gas temperature for an initially uniform medium with gas and radiation in thermal equilibrium as a function of optical depth $\tau$ and gas to radiation pressure ratio $\beta_{\rm r}$. We used 64 zones in 1D and 64 samples per zone. Noise is generated from our method for calculating the radiation pressure tensor. 
Noise in the gas temperature grows only slowly once $\beta_{\rm r} \lesssim 1$, rather than rapidly going unbounded as in a traditional explicit Monte Carlo method. This is due to our semi-implicit update that drives the gas and radiation towards thermal equilibrium. Additionally, in thermal equilibrium $u_{\rm r} \sim T_{\rm r}^4$, and noise in $\tensor{\pi}{^{(i)}_{(j)}}$ directly affects $u_{\rm r}$ rather than $T_{\rm r}$. Note that even with only 64 samples, maximum noise in the gas temperature is of order 0.5\% (noise in the radiation temperature is generally $\sim 10^{-4}$.}
\label{fig:sine_diffusion_err}
\end{figure}

We parameterize the stability by $\langle | T_{\rm g}- T_0| \rangle/T_0$, where $T_{\rm g}$ is the gas temperature and $\langle \cdot \rangle$ is the average over the domain. We set $\beta_{\rm r}$ and $\tau$ by setting the gas density and the gray opacity $\kappa$. We run each realization for $t = 10 L/c$, which appears to lead to saturated gas temperature errors. We consider a range of optical depths and gas-to-radiation pressures, $\tau \in [10^{-2}, 10^2]$, and $\beta_{\rm r} \in [10^{-5}, 10^2]$. The mean error at final time for each of these realizations is shown in Figure \ref{fig:sine_diffusion_err}. Evidently our method is stable for every combination of parameters we consider, and mean gas temperature errors remain good $(\lesssim 0.5\%)$ even in the most extreme cases.

\subsection{Radiation Magnetohydrodynamic Linear Modes}

We revisit relativistic radiation MHD linear modes (\citealt{Jiang+2012}, \citealt{Sadowski+2014}, \citealt{Ryan+2015}). These are derived assuming the Eddington closure, and so we focus on the optically thick regime here. Solutions are generated with a symbolic linear modes package (\citealt{Chandra+2017}). We specialize to optically thick radiation-modified fast magnetosonic modes at different gas to radiation pressures $\beta_{\rm r}$. We do not refine or derefine samples for this test. We construct eigenmodes of the form $\delta \sim \exp\left( \omega t + i k x\right)$ for variation in ${\bf P} =(\rho_0 + \delta \rho$, $u_{{\rm g},0} + \delta u_{\rm g}$, $\delta u^1$, $\delta u^2$, $B^1_0$, $B^2_0 + \delta B^2$, $u_{{\rm r},0} + \delta u_{\rm r}$, $\delta F^1$, $\delta F^2 )$. Optical depth per wavelength $\tau = 20$, divided evenly between gray absorption opacity $\kappa^{\rm a}$ and gray scattering opacity $\kappa^{\rm s}$, for wavenumber $k = 2 \pi$. Modes are simulated with an amplitude $\delta = 10^{-4}$. The background equilibrium is $\rho_0 = 1$, $u^1_0 = u^2_0 = F^1_0 = F^2_0 = 0$, $E_0 = a_{\rm r} (P_0/\rho_0)^4$ and $u_{{\rm g},0}$ and $B^1_0 = B^2_0$ are determined by plasma $\beta_{\rm m} = 1$ and gas to radiation pressure $\beta_{\rm r}=(10,1,0.1)$. The adiabatic index $\gamma =5/3$. Units are such that $k_{\rm B} = c = a_{\rm r} = 1$. The three modes we consider are given in Table \ref{tab:modes}.

\begin{figure*}
\includegraphics[width=\textwidth]{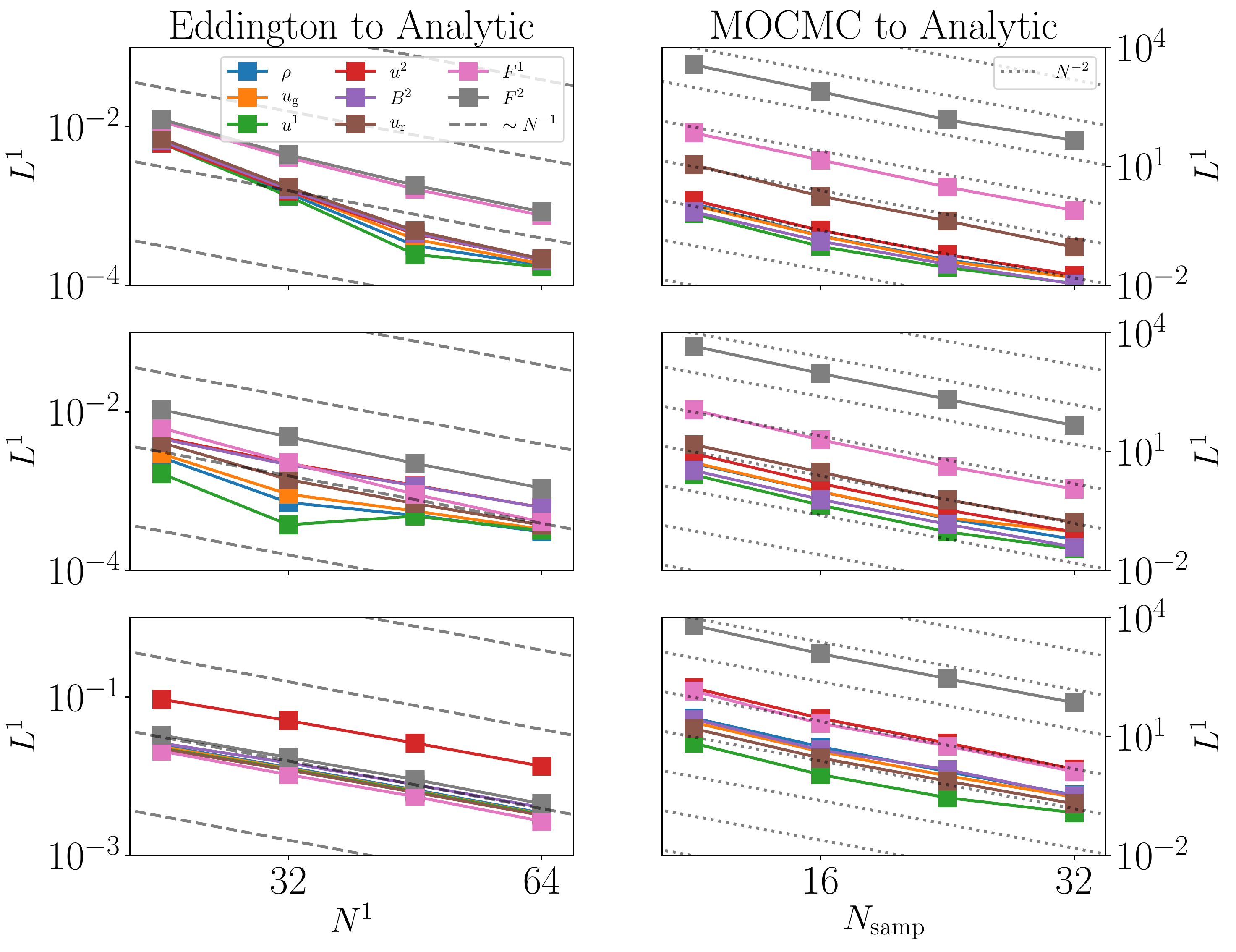}
\caption{Convergence of radiation-modified fast magnetosonic modes for, from top to bottom, $\beta{\rm r} = (10, 1, 0.1)$. Due to the two independent resolution parameters in MOCMC, we first study convergence in number of grid zones $N^1$ of the numerical solution using the Eddington closure with respect to the analytic solution, and then convergence in number of samples per zone $N_{\rm samp}$ of the numerical MOCMC solution at fixed $N^1=256$ with respect to the same analytic solution. Different gas to radiation pressure ratios $\beta_{\rm r}$ are shown; the plasma $\beta_{\rm m} = 1$. All modes shown are optically thick, $\tau = 20$. Here, the $L^1$ norm corresponds to the fractional error relative to mode amplitude in each zone. For the Eddington closure, when advection errors dominate we expect second order convergence in $N^1$; when coupling dominates, we expect first order convergence in $N^1$. The MOCMC solution converges as $N_{\rm samp}^{-2}$, indicating a second order-accurate integration of the Eddington tensor, and negligible truncation error in the sample updates themselves.}
\label{fig:linmode_edd_conv}
\end{figure*}

Each mode is simulated for one wave crossing time, $2 \pi / | {\rm Imag} \left( \omega\right)|$. We study convergence. However, we have two resolution parameters: number of zones $N^1$, and number of samples per zone $N_{\rm samp}$, which introduce similar errors. We therefore approach convergence in two steps. First, we study convergence of the MOCMC code with the Eddington closure relative to the analytic solution with increasing $N^1$ (this subsystem has no $N_{\rm samp}$ dependence). Figure \ref{fig:linmode_edd_conv} shows the expected $(N^1)^{-1}$ convergence in all variables at large $N^1$ (these modes are optically thick, and our source term evaluations are first order accurate in time. Next, we fix $N^1 = 256$, and reintroduce samples for computing the pressure tensor. We then study convergence of the full MOCMC solution with increasing $N_{\rm samp}$ relative to the analytic solution. When the truncation error is dominated by the integration of the Eddington tensor, MOCMC converges as $\sim N_{\rm samp}^{-2}$.

\begin{deluxetable*}{cccc}
\tablecaption{ Radiation-Modified Fast Magnetosonic Modes \label{table:linearmodes}} 
\tablehead{ & $ \beta_{\rm r} = 10$& $\beta_{\rm r} = 1$& $\beta_{\rm r} = 0.1$} 
\startdata 
$\omega$ & $-0.0244360 + 0.896566i$ & -0.0932061-0.983571i &-0.163372-1.79706i \\
\hline
$\delta \rho/\delta$  &$ 0.981029$ & $0.980516 $& $0.926746  $\\ 
$\delta u_{\rm g} /\delta$ & $0.0146350+0.00156849i $& $0.0127094-0.00173903i $&$ 0.0122892-0.000860716i$\\ 
$\delta u^1 /\delta$ & $-0.139986-0.00381533i $&$ 0.153490-0.0145452i$ &$ 0.265060-0.0240967i $\\ 
$\delta u^2/\delta $ & $0.0640717-0.00344325i $&$ -0.0508806-0.00831436i$ &$ -0.0162519-0.00161307i$\\ 
$\delta B^2 /\delta$ & $0.116682-0.00296728i $&$ 0.105954+0.00679059i$&$0.0802285+0.000875043i $\\ 
$\delta u_{\rm r}/\delta $ & $0.00372334+0.00109715i $&$0.0230965-0.0135388i$ & $0.241633-0.0682760i$\\ 
$\delta F^1 /\delta$ & $9.71549\times10^{-5}-0.000378002i$& $-0.00127282-0.00229245i$&$-0.00427347-0.0195488i $\\ 
$\delta F^2 /\delta$ & $-5.46703\times10^{-7}-7.65533\times10^{-6}i$ & $7.94614\times10^{-6}-6.76842\times10^{-5}i$& $3.88462\times10^{-5}-0.000392639i$\\ 
\enddata 
\tablecomments{Eigenmodes of the form $\delta \sim \exp\left( \omega t + i k x\right)$ for the equations of covariant radiation hydrodynamics in Minkowski space with the Eddington closure. Optical depth per wavelength $\tau = 20$ and plasma $\beta_{\rm m} = 1$. Our simulations use an amplitude $\delta = 10^{-4}$.}
\label{tab:modes}
\end{deluxetable*}

\subsection{Relativistic Nonlinear Waves}

\cite{Farris+2008} introduced a method for calculating 1D relativistic radiation hydrodynamic waves in flat spacetime with an assumed Eddington closure and a gray absorption opacity, along with four example solutions that have frequently been reproduced with relativistic radiation hydrodynamics codes (\citealt{Fragile+2012}, \citealt{Roedig+2012}, \citealt{Sadowski+2013}, \citealt{McKinney+2014}, \citealt{Ryan+2015}). These four example solutions are (Case 1) a gas pressure-dominated nonrelativistic strong shock (Case 2) a gas pressure-dominated mildly relativistic strong shock (Case 3) a gas pressure-dominated highly-relativistic wave (Case 4) a radiation pressure-dominated, mildly relativistic wave. We adopt the parameters from \cite{Farris+2008}. \cite{Ryan+2015} simulated Cases 1, 2, and 3 with full transport. See also \cite{OhsugaTakahashi2016} for another full transport method applied to this problem.

We initialize these problems as shocktubes and allow them to evolve to equilibrium inside the code. We perform these simulations with both the Eddington closure and the full MOCMC machinery. The Eddington closure provides the reference solution and tests part of our numerical framework against the analytic solution (\citealt{Farris+2008}); the MOCMC solution, being full transport, will not agree with the analytic solution on scales of an optical depth, which in all cases is approximately the scale of the interface structure. 

\begin{figure}
\includegraphics[width=\columnwidth]{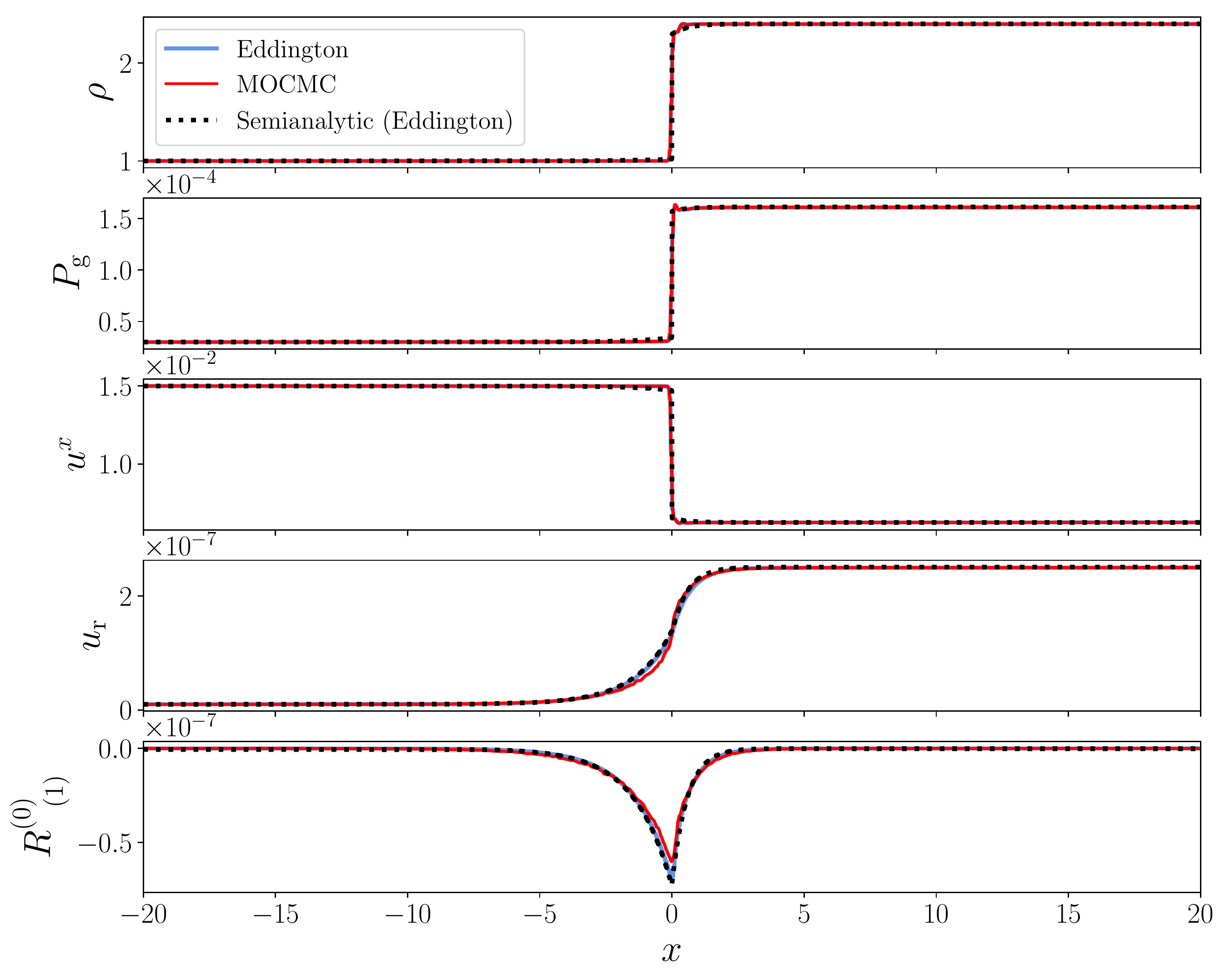}
\caption{A nonrelativistic, gas pressured-dominated, weak shock initialized as a shock tube and evolved for $t = 40$. The radiation has little effect on the fluid, and this is largely a test of radiation transport over finite optical depth in a nonuniform medium.}
\label{fig:farris1}
\end{figure}

\begin{figure}
\includegraphics[width=\columnwidth]{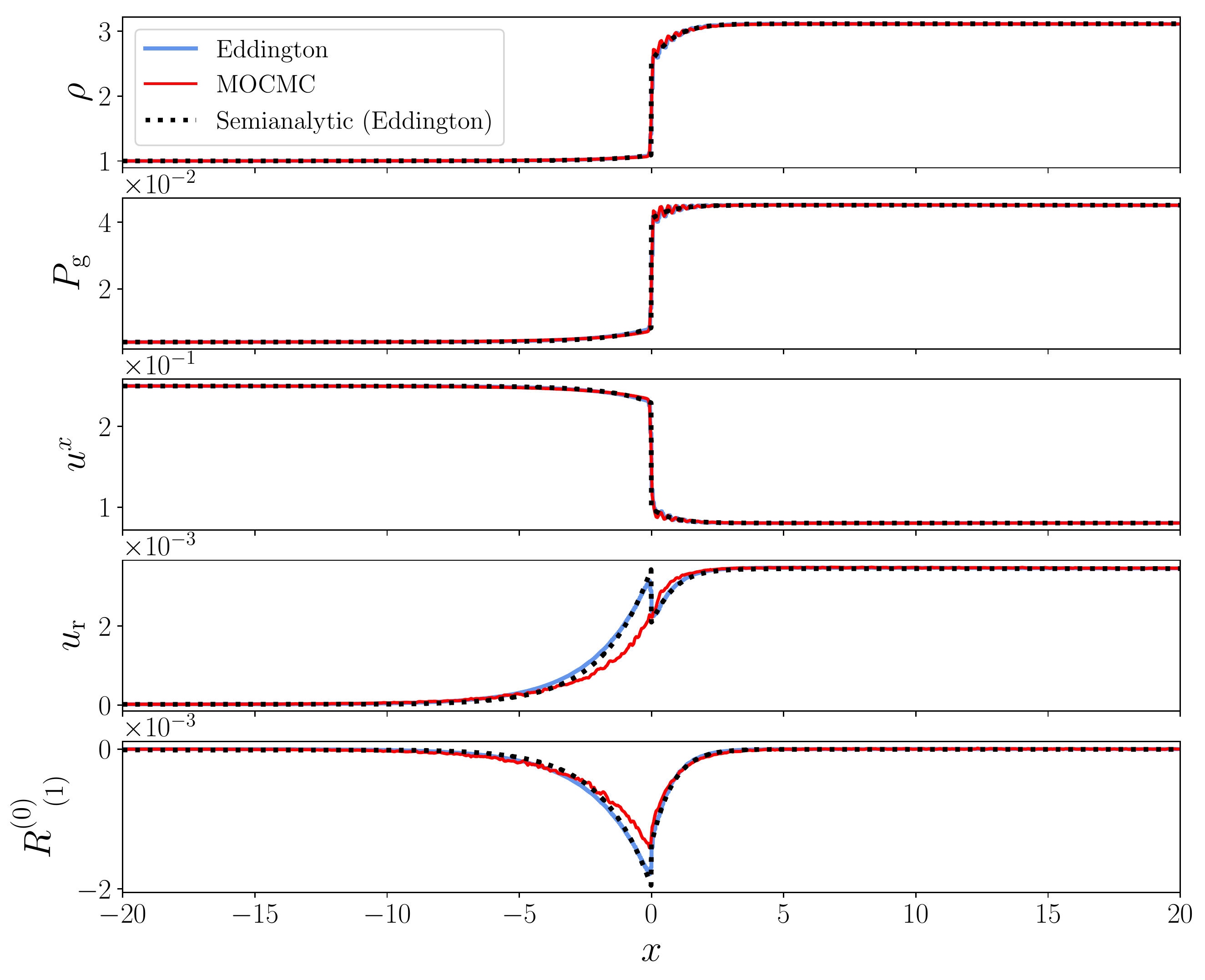}
\caption{A mildly relativistic, gas pressure-dominated, strong shock at $t = 400$. The transport solution is qualitatively dissimilar from the Eddington approximation, which produces large discontinuities in the comoving radiation energy density and flux. Note the good agreement between MOCMC and the solution given by the explicit Monte Carlo method \bhlight{} in \citealt{Ryan+2015}).}
\label{fig:farris2}
\end{figure}

\begin{figure}
\includegraphics[width=\columnwidth]{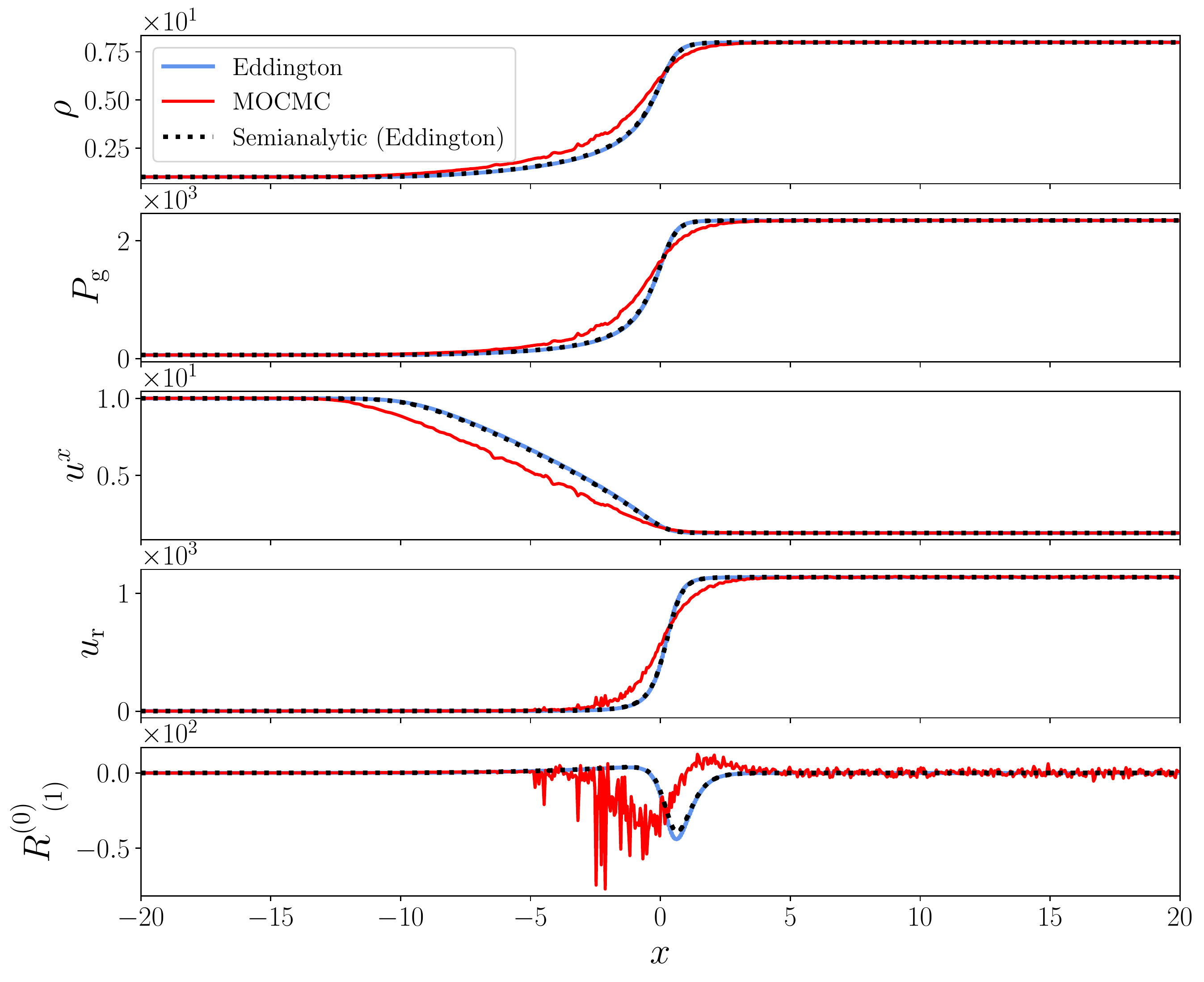}
\caption{The highly relativistic relativistic gas pressure-dominated wave at $t = 80$. This test exposes a significant pathology in MOCMC due to our sampling method. Initially, samples are distributed uniformly in the comoving frame of the fluid in each zone. However, in this test these samples quickly pass from a $\gamma \sim 10$ region to a $\gamma \sim 1$ region. As a result, most of the samples downstream of the interface will be almost colinear in the fluid frame, leading to a challenging reconstruction operation, and requiring {\it in situ} resampling.}
\label{fig:farris3}
\end{figure}

\begin{figure}
\includegraphics[width=\columnwidth]{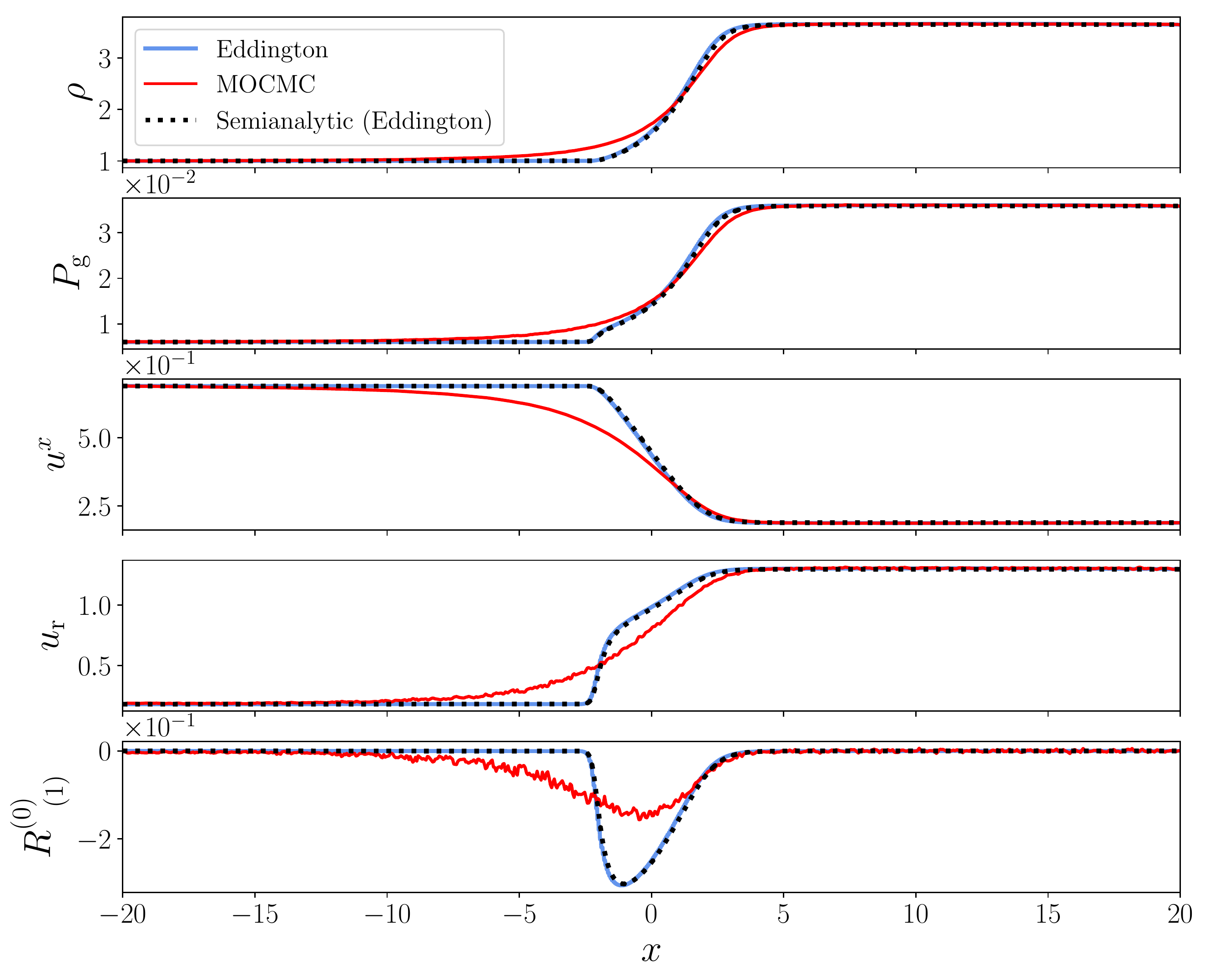}
\caption{The radiation pressure-dominated, mildly relativistic wave at $t = 150$. Despite $\beta_R \sim 0.03 \ll 1$, our semi-implicit MOCMC method is stable despite solving the transport equation using particles. The sharp features in the comoving radiation energy density and radiation flux in the Eddington approximation are not present in the transport solution. The explicit Monte Carlo method \bhlight{} (\citealt{Ryan+2015}) was not able to stably evolve this configuration.}
\label{fig:farris4}
\end{figure}

For all cases, we use 800 spatial zones and, for the MOCMC solution, approximately 64 samples per zone after refinement. Each sample carries 50 frequency bins. The left and right initial interface states are enforced at the boundary in the fluid and radiation moment variables, while the pressure tensor is taken to be Eddington and the samples are thermal and uniformly distributed in solid angle in the comoving frame. The four waves are shown, respectively, in Figures \ref{fig:farris1}, \ref{fig:farris2}, \ref{fig:farris3}, and \ref{fig:farris4}. The particle noise in the MOCMC representation of the Eddington tensor is mostly small and in all cases does not prevent the code from being stable or accurate, even when radiation pressure is dominant. The most significant pathology is when the Lorentz factor changes dramatically across the wave (Case 3); the resulting beaming of samples leads to poor sampling of solid angle in the comoving frame. The code then relies on the resampling procedure for resolving the pressure tensor.


\subsection{Novikov-Thorne Hohlraum}

We now consider radiation transport in curved spacetime. We essentially repeat the 2D flat space hohlraum test in the Schwarzschild geometry for a $1~M_{\odot}$ black hole, now with the radiating boundary condition a thin disk at the midplane from the innermost stable circular orbit to $10~GM/c^2$. 
The disk has the temperature profile of a thin disk (\citealt{NovikovThorne1973}) with anomalous viscosity $\alpha=0.05$. 
The disk and atmosphere are static in the normal observer frame. 
This disk radiates into vacuum; at finite time, we measure the radiation energy density in the normal observer frame. 

To construct a time-dependent solution to the equation of radiative transfer, we extend the procedure from \ref{sec:2d_hohlraum} to a geometrically thin, optically thick disk in the Schwarzschild geometry. The procedure is essentially unchanged except now, instead of propagating rays backwards in time along straight lines in flat space, we sample geodesics uniformly in the normal observer frame (e.g.\ \citealt{Bardeen+1972}) and propagate them along geodesics until they either exit the outer radial boundary, cross the event horizon, or intersect with the thin disk. For geodesics that intersect the disk, we set their invariant intensity to the Planck function at the disk temperature, and integrate this over frequency in the originating normal observer frame to get the contribution to radiation energy density $u_{\rm r,normal}$ in that frame.

\begin{figure}
\includegraphics[width=\columnwidth]{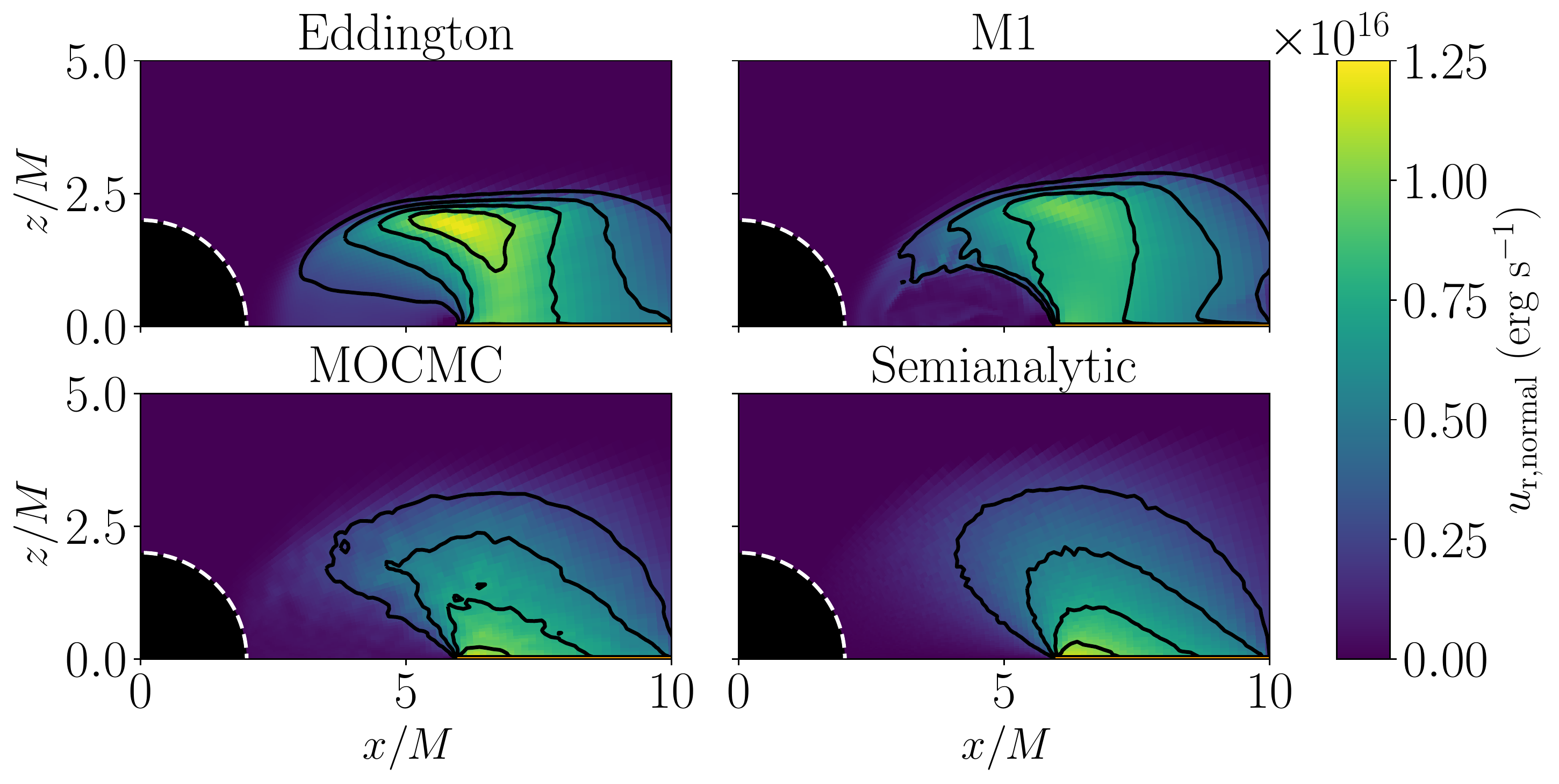}
\caption{Time-dependent transport test in the Schwarzschild spacetime for a radiating disk at $t = 5M$. Radiation energy density in the normal observer frame is shown. The radiating disk is at the midplane from $r=6 GM/c^2$ to $r=10 GM/c^2$, and the black circles denotes the event horizon of the black hole.  Eddington, M1, and MOCMC (with 64 samples per zone) closures are shown against the semianalytic solution. The Eddington and M1 closures produce similar results, with large radiation energy densities far from the disk at finite time, and, particularly in the case of M1, a sharp boundary to a vacuum region in the midplane inside the innermost stable circular orbit. As in flat spacetime, at finite time the MOCMC solution corresponds much more closely to the semianalytic solution than either of the moment closures.}
\label{fig:nttrans}
\end{figure}

Figure \ref{fig:nttrans} shows the results of this test, both for the full MOCMC code as well as Eddington and M1 closures, alongside the semianalytic solution. For the simulations, we adopt axisymmetry in modified Kerr-Schild (\citealt{Gammie+2003}) coordinates with refinement parameter $h=0.3$ and use $128\times129$ zones in $X^1$ and $X^2$, the odd number of zones in $X^2$ for symmetry about the midplane. The lower hemisphere is not shown. We set the outer radius at $20 GM/c^2$. The thermal radiating disk boundary (we adopt a hemispheric approach similar to the 2D hohlraum test in Section \ref{sec:2d_hohlraum}) is enforced at the midplane every substep, both in the radiation moments and in the samples (i.e.\ the radiating disk is infinitely optically thick in our implementation).

\subsection{Isothermal Schwarzschild Atmosphere}

We repeat the isothermal pressure-supported atmosphere test in the Schwarzschild geometry close to the event horizon from \cite{Ryan+2015}. Reflecting spherical shells are placed at inner and outer radii $r_{\rm in} = 3.5 GM/c^2$ and $r_{\rm out} = 20 GM/c^2$, and gas between these shells is allowed to reach radiative equilibrium through a gray opacity $\kappa$. The radiation acts like a heat conduction, leading to a temperature profile that is isothermal modulo a redshift factor,
\begin{align}
T\left(r \right) = T_{\infty}/\sqrt{-g_{00}}
\end{align}
where $T_{\infty}$ is the temperature at large radius. The solution is determined by this temperature profile and mechanical equilibrium, $\tensor{T}{^{\mu r}_{;\mu}} = 0$. For a $\gamma$-law equation of state, the solution is evaluated by solving
\begin{align}
\frac{dP}{dR} = -\frac{\left(\rho + \frac{\gamma}{\gamma - 1}P_{\rm g} \right)}{r^2\left( 1 - \frac{2}{r}\right)}.
\end{align}

\begin{figure}
\includegraphics[width=\columnwidth]{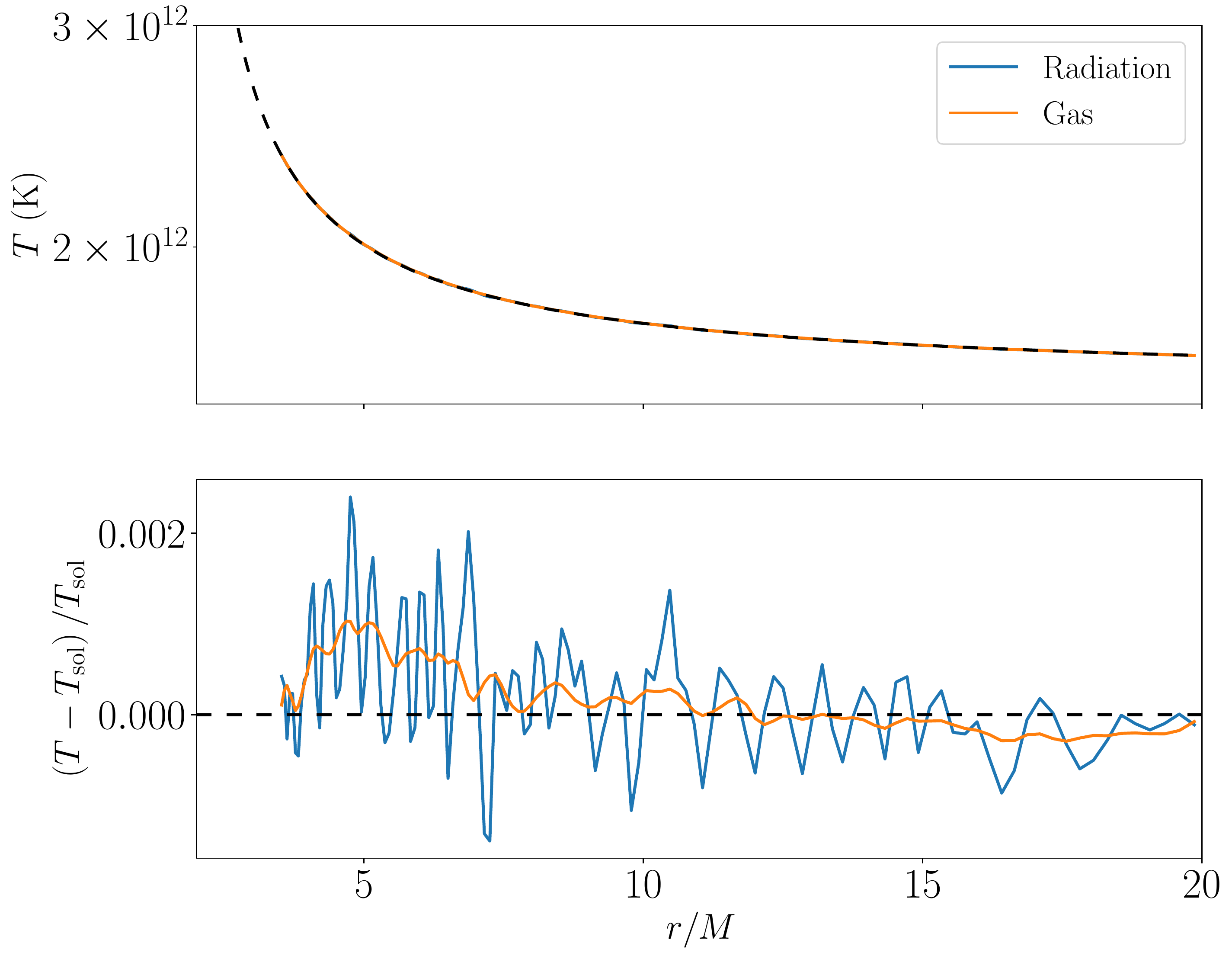}
\caption{Redshifted isothermal Schwarzschild atmosphere at $t=500 GM/c^3$. The temperature profiles of both the radiation and the gas are shown, as well as the residuals relative to the semianalytic solution. $I_{\nu}/\nu^3 = B_{\nu}/\nu^3 = {\rm const}$ everywhere in the domain. The solution is enforced in the ghost zones at both radial boundaries. At least some of the structure in the residuals may be due to our treatment of the boundary conditions.}
\label{fig:bhtherm}
\end{figure}

Because all rays propagate back to $t = - \infty$, $I_{\nu}/\nu^3 = B_{\nu}/\nu^3$ everywhere in the domain and this is an exact solution of the equations of radiation hydrodynamics with full transport in curved spacetime. This problem can be cast in terms of three dimensionless parameters: (1) the ratio of the inner atmospheric scaleheight $H$ to $r_{\rm in}$, $H/r_{\rm in} = k_{\rm B} T_{\rm in} r_{\rm in}/(\mu m_{\rm p} G M)$ where $\mu$ is the mean molecular weight (2) the ratio of gas to radiation pressure at the inner boundary, $\beta_{\rm r} = \mu m_{\rm p} a_{\rm r} T_{\rm in}^3/(3 \rho_{\rm in} k_{\rm B})$ (3) the optical depth across the domain $\tau = \kappa \rho_{\rm in} (r_{\rm out} - r_{\rm in})$. We set the black hole mass $M = M_{\odot}$, $H/r_{\rm in} =1.60$, $\beta_{\rm r} = 43.5$, and $\tau = 5$. We set up the simulation in 1D, with 128 grid zones. The exact solution is enforced at the boundaries. We run for $t = 500 GM/c^3$. The result is shown in Figure \ref{fig:bhtherm}.

\section{Conclusions}
\label{sec:conclusion}

We have presented a numerical method for covariant radiation magnetohydrodynamics with frequency-dependent transport that is stable, accurate, and efficient for a wide range of optical depths and radiation pressures relevant to the black hole accretion problem. The essential novelty is the discretization of the radiation field. Specific intensities are transported along characteristics, and the radiation distribution function in fluid zones is reconstructed by the set of samples in each zone at each timestep. Source terms are evaluated in a deterministic, fully nonlinear, and implicit fashion, avoiding difficulties encountered by Monte Carlo methods for large optical depths and/or short interaction timescales. This solution to the transport equation is used to close a set of moment equations, providing a numerical solution to the full transport equation. The continuous nature of our method also means that the radiation field, and radiation interactions, are generally less noisy than in Monte Carlo; in particular, errors decrease with number of samples at least as $N_{\rm samp}^{-1}$, rather than the canonical Monte Carlo $N_{\rm samp}^{-1/2}$. By transporting an array of intensities at different frequencies along a common geodesic, we significantly reduce the algorithmic complexity of the transport operator in multigroup problems. Our treatment, in which specific intensity samples can be readily re-sampled locally, is also advantageous for large dynamic spatial ranges (such as logarithmic grids in simulations of thick disks). This should lead to improved load balancing and could also be a benefit in future numerical methods with adaptive mesh refinement. 

The method we have presented is a particular realization of a class of methods, in which integrations in solid angle over long characteristics are used to evaluate unknowns in the continuum evolution of the radiation four-momentum. In particular, one could adopt different integration methods, like a simple sum (which would lead to Monte Carlo-like $N_{\rm samp}^{-1/2}$ errors) or fitting spherical harmonics to the set of samples, which could lead to higher angular and spatial accuracy.

\acknowledgments
It is a pleasure to thank C.\ F.\ Gammie, Y.-F.\ Jiang, J.\ Miller, D.\ Radice, M.\ Chandra, C.\ White, F.\ Foucart, J.\ Stone, R.\ Wollaeger,
and the members of the {\tt horizon} collaboration\footnote{\url{http://horizon.astro.illinois.edu/}} for valuable discussions. This work was supported by the US Department of Energy through the Los Alamos National Laboratory. Los Alamos National Laboratory is operated by Triad National Security, LLC, for the National Nuclear Security Administration of U.S. Department of Energy (Contract No. 89233218CNA000001). This work was supported by the Laboratory Directed Research and Development program of Los Alamos National Laboratory under project numbers 20170527ECR and 20180716PRD2. This work has been assigned a document release number LA-UR-19-24802.

\end{document}